\documentclass[aps,prx,floatfix,twocolumn,longbibliography,nofootinbib]{revtex4-2}

\usepackage[utf8]{inputenc}
\usepackage[T1]{fontenc}
\usepackage{physics}
\usepackage{graphicx}
\usepackage{amsmath, amsfonts, amssymb}
\usepackage[breaklinks=true, colorlinks]{hyperref}
\hypersetup{linkcolor=red, citecolor=blue}

\usepackage{bbm, bm}
\usepackage[dvipsnames]{xcolor}
\usepackage{makecell}
\usepackage{upgreek}
\usepackage{nicefrac}

\newcommand{\OO}{\mathcal{O}}
\newcommand{\ttheta}{\bm{\theta}}
\newcommand{\ppi}{\bm{\pi}}
\newcommand{\avg}[1]{{\left\langle {#1} \right\rangle }}
\newcommand{\Var}[0]{{ \text{Var}  }}
\newcommand{\diag}[0]{{ \text{diag} }}

\DeclareMathOperator*{\argmin}{argmin}

\begin{document}

\title{Variational quantum dynamics of two-dimensional rotor models}

\author{Matija Medvidović}
\affiliation{Center for Computational Quantum Physics, Flatiron Institute, 162 5th Avenue, New York, New York 10010, USA}
\affiliation{Department of Physics, Columbia University, New York 10027, USA}

\author{Dries Sels}
\affiliation{Center for Computational Quantum Physics, Flatiron Institute, 162 5th Avenue, New York, New York 10010, USA}
\affiliation{Center for Quantum Phenomena, Department of Physics, New York University, 726 Broadway, New York, NY, 10003, USA}

\date{\today}

\begin{abstract}
    We present a numerical method to simulate the dynamics of continuous-variable quantum many-body systems. Our approach is based on custom neural-network many-body quantum states. We focus on dynamics of two-dimensional quantum rotors and simulate large experimentally relevant system sizes by representing a trial state in a continuous basis and using state-of-the-art sampling approaches based on Hamiltonian Monte Carlo. We demonstrate the method can access quantities like the return probability and vorticity oscillations after a quantum quench in two-dimensional systems of up to 64 (8 $\times$ 8) coupled rotors. Our approach can be used for accurate nonequilibrium simulations of continuous systems at previously unexplored system sizes and evolution times, bridging the gap between simulation and experiment.
\end{abstract}

\maketitle

\section{Introduction}
\label{sec:introduction}

Nonequilibrium quantum many-body physics has been at the forefront of condensed matter, atomic physics and chemistry research for over a decade~\cite{rabitz93, rmpPolkovnikov11}. The field is driven by remarkable progress in our ability to coherently control matter at the atomic scale. This control has resulted in the creation of novel phases of matter, including observations of light-induced superconductivity~\cite{cavalleri21}, cavity-enhanced chemical reactions~\cite{Ebbesen19} and dynamical phase transitions~\cite{Zhang17}.

The capacity to precisely control~\cite{Sivak2022, Porotti2022, Metz2022, Bukov2018} modern quantum experiments and hardware is becoming increasingly limited by numerical simulation of the real-time evolution of quantum systems. At its core, the problem is related to fast entanglement growth in systems out of equilibrium, which forces one to keep track of all the intricate correlations that build up in the system. While there has been considerable progress~\cite{zaletel15, wurtz18, Czarnik2019, hubig20, zou20}, challenges remain, in particular if one moves away from one-dimensional spin models. 

Recently, it has been proposed that methods inspired by classical and quantum machine learning might alleviate some of these problems~\cite{Carleo2012, Carleo2017_2, Schmitt2020, Hofmann2022, Barison2021}. In practice, however, it has been difficult to achieve reliable results due numerical instabilities resulting from a combination of Monte Carlo noise and flatness of the quantum geometry of modern neural-network wave functions~\cite{Czischek18, Schmitt2020, amari_natural_1998, Yuan2019, stokes_quantum_2020}.

In this work, we present an approach for capturing long-time dynamics of two-dimensional (2D) lattice models with continuous degrees of freedom, using a combination of methods that were previously unexplored in the field of variational simulations -- the Hamiltonian Monte Carlo sampler, a tailored variational ansatz and proper regularization of the projected dynamics. We focus on the quantum rotor model with direct applications to arrays of coupled Josephson junctions and explore previously unreachable system sizes and evolution times, up to $8\times 8$ square lattices.

The paper is organized as follows. First, we introduce the physics of the quantum rotor model and the variational wavefunction. Then, we outline the Hamiltonian Monte Carlo sampler and its connection to the time-dependent variational Monte Carlo algorithm. Finally, we present results for the two-dimensional model, showing magnetization, vorticity and the Loschmidt echo converging to appropriate equilibrium values. Our Monte Carlo results are substantiated by self-consistency checks when key hyperparameters are changed and by comparing our approach to tensor-network calculations in one and two spatial dimensions.

\begin{figure*}[!t]
    \centering
    \includegraphics[width=\textwidth]{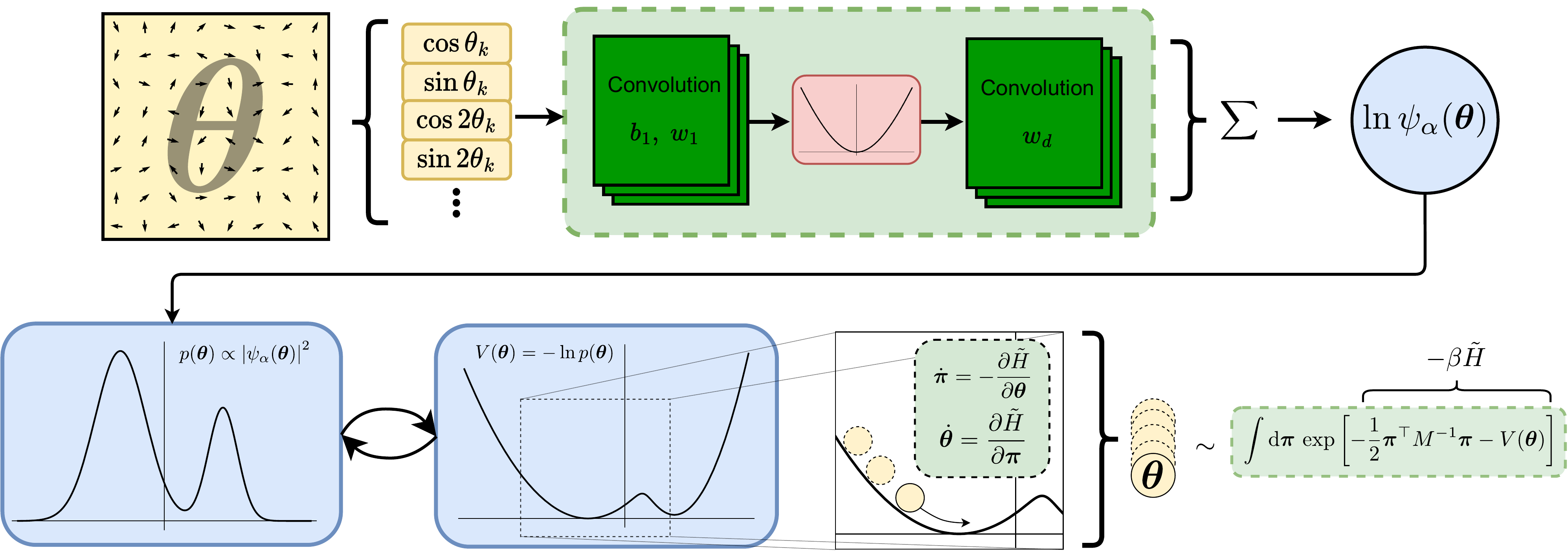}
    \caption{
        \textbf{Top}:
        The ansatz $\psi _\alpha (\ttheta)$ architecture used for simulations of two-dimensional QRM systems. It amounts to a two-layer convolutional neural network with an activation function given by Eq.~\ref{eq:taylor_activation}. To enforce periodicity and improve expressivity, we precalculate sines and cosines of input angles which are treated as different input channels by the CNN. The final layer outputs a single channel and all of its components are summed into a single complex number (because of complex parameters $\alpha \in \mathbb{C}^P$) we then interpret as $\ln \psi _\alpha (\ttheta)$.
        \textbf{Bottom}:
        An illustration of the Hamiltonian Monte Carlo algorithm. Dummy momentum variables are introduced and sampling the given $N$-dimensional probability distribution is rewritten in $2N$-dimensional phase space with an artificial effective Hamiltonian $\Tilde{H}$. Samples are collected as snapshots of solutions of Hamilton's equations of motion.
    }
    \label{fig:diagram}
\end{figure*}

\section{Model and Methods}
\label{sec:methods}

Consider a system of continuous planar rotors, whose angles $\theta _k$ (with respect to an arbitrary axis) could, for example, represent superconducting phases of adjacent Josephson junctions on a lattice $\Lambda$ with $N$ sites. We use the basis $\ket{\ttheta} \equiv \ket{\theta _1, \ldots, \theta _N}$ for the Hilbert space $\mathcal{H}$. We start with an effective Hamiltonian that captures the relevant physics of superconducting Josephson junctions \cite{Vogt2015, Martinoli2000, Kockum2019}:
\begin{equation}
    \label{eq:hamiltonian}
    H = \frac{g J}{2} \sum _k L_k ^2 - J \sum _{\langle k, l \rangle} \hat{\vb{n}} _k \cdot \hat{\vb{n}} _l \; ,
\end{equation}

\noindent where $ L_k = -i \, \partial _{\theta _k}$ and $\hat{\vb{n}} _k = (\cos \theta _k, \sin \theta _k )$ in the continuous basis $\ket{\ttheta}$ of choice. The Hamiltonian in Eq.~\ref{eq:hamiltonian} is often called the quantum rotor model (QRM). Its equilibrium properties \cite{Jose1977} have been studied using variational Monte Carlo (VMC) \cite{Stokes2021} and other quantum Monte Carlo (QMC) \cite{Jiang2019} methods. Perhaps most notably, the quantum critical point separating the disordered and O(2) broken phase has been predicted at $g_c \approx 4.25$.

However, as noted in the introduction, real-time evolution properties of the QRM have barely been explored. This is mainly due to the lack of suitable methods that can access experimentally relevant times $t \gg J^{-1}$, at large system sizes in two dimensions. The ability to simulate relatively large system sizes is not only of theoretical interests but has technological applications in the study of dynamics of arrays of coupled Josephson junctions~\cite{trebst22}.

The evolution equation for a state $\Psi = \Psi (\ttheta) $, in the continuous basis $\ket{\ttheta}$, reads
\begin{equation}
    \label{eq:pde}
    i \frac{\partial \Psi }{\partial t} = - \frac{g J}{2} \sum _k \frac{\partial ^2 \Psi}{\partial \theta ^2 _k} - J \sum _{\langle k, l \rangle} \cos ( \theta _k - \theta _l ) \Psi
\end{equation}

\noindent with appropriate periodic boundary conditions
\begin{equation}
    \Psi (\theta_1, \ldots, \theta _k + 2 \pi, \ldots, \theta _N) = \Psi (\theta_1, \ldots, \theta _k, \ldots, \theta _N)
\end{equation}

\noindent for each rotor $k$. Eq.~\ref{eq:pde} is prohibitively expensive to solve exactly even for a handful of interacting rotors. The continuous nature of the $\ket{\ttheta}$ basis exacerbates the problem.

\subsection{Variational simulation}
\label{sec:variational_sim}

We represent a quantum state using a wavefunction $\psi _\alpha (\ttheta) $ where $\alpha \in \mathbb{C} ^P$ is a set of $P$ real or complex variational parameters. Since any $\ket{\psi} \in \mathcal{H}$ admits an expansion in terms of $\ket{\ttheta}$, we define the following un-normalized variational quantum state (VQS):
\begin{equation}
    \label{eq:vqs}
    \ket{\psi _\alpha } = \int \dd \ttheta \; \psi _\alpha (\ttheta) \ket{\ttheta}
\end{equation}

\noindent where $\dd \ttheta \equiv \dd \theta _1 \cdots \dd \theta _N$. The integral is performed over the cube $[-\pi,\pi]^N$.

Building on previous work on continuous systems \cite{Carleo2017_2}, our simulation of the real-time dynamics of the state given in Eq.~\ref{eq:vqs} is based on the time-dependent variational Monte Carlo (t-VMC) method~\cite{Becca2017, Carleo2012}. The core assumption that allows us to approximately solve Eq.~\ref{eq:pde} is that of time dependence of parameters $\alpha = \alpha (t)$.

Optimal trajectories $\alpha (t)$ induced by unitary Hamiltonian evolution $e^{-i H t} \ket{\psi _\alpha}$ can conveniently be found by extremizing the time-dependent variational principle (TDVP)~\cite{Yuan2019} action
\begin{equation}
    \label{eq:tdvp}
    \mathcal{C}[\alpha] = \int \dd t \bra{\Psi _{\alpha (t)}} \left( i \frac{\dd}{\dd t} - H \right) \ket{\Psi _{\alpha (t)}} \; .
\end{equation}

\noindent where $\ket{\Psi _\alpha}$ is a normalized version of state $\ket{\psi _\alpha}$. Optimal evolution equations read $i \, S \dot{\alpha} = g$, where
\begin{equation}
    \label{eq:averages}
    \begin{gathered}
        S _{\mu \nu} = \avg{\OO ^\dagger _\mu \OO _\nu} - \avg{\OO ^\dagger _\mu} \avg{\OO _\nu \vphantom{\OO ^\dagger _\mu} } \\
        g _\mu = \avg{\OO ^\dagger _\mu H} - \avg{\OO ^\dagger _\mu} \avg{H \vphantom{\OO ^\dagger _\mu} } \;
    \end{gathered}
\end{equation}

\noindent with averages $\avg{\cdot} \equiv \nicefrac{\bra{\psi _\alpha} \cdot \ket{\psi _\alpha} }{ \braket{\psi _\alpha} }$ being performed at time $t$ (i.e. for $\alpha = \alpha (t)$). Operator $\OO _\mu$ is defined by $\partial _{\alpha _\mu} \ket{\psi _\alpha} = \OO _\mu \ket{\psi _\alpha}$. We note that the matrix $S$ is commonly called the \textit{quantum geometric tensor} (QGT)~\cite{Sorella1998, amari_natural_1998, stokes_quantum_2020} and corresponds to the metric tensor of the parameter manifold induced by the distance in $\mathcal{H}$ between un-normalized states defined in Eq.~\ref{eq:vqs}. In Eqs.~\ref{eq:averages}, we have chosen our ansatz $\psi _\alpha$ such that it is a holomorphic function of complex parameters $\alpha$.

Since quantum averages over an exponentially large Hilbert space $\mathcal{H}$ in the TDVP Eq.~\ref{eq:averages} cannot be computed exactly, Markov chain Monte Carlo (MCMC) sampling methods are often employed \cite{Metropolis1953, Hastings1970}. In VMC calculations, it is common to rewrite quantum averages, such as those in Eq.~\ref{eq:averages}, as expressions amenable to estimation through sampling. For example, in the case of the Hamiltonian $H$, we obtain the \textit{local energy} $E_L$:
\begin{equation}
\label{eq:local_energy}
    \avg{H} = \frac{\bra{\psi _\alpha} H \ket{\psi _\alpha}}{\braket{\psi _\alpha}} = \int \dd \ttheta \; p _\alpha (\ttheta) \, E_L (\ttheta )
\end{equation}

\noindent where
\begin{equation}
    p_\alpha (\ttheta) \propto \left| \psi _\alpha (\ttheta) \right| ^2 \quad \text{and} \quad E_L (\ttheta) = \frac{\bra{\ttheta} H \ket{\psi _\alpha}}{\braket{\ttheta}{\psi _\alpha}} \; .
\end{equation}

For more details about the specific sampling algorithm employed in this work, we refer the reader to Sec~\ref{sec:hmc} and Appendix~\ref{appendix:hmc}.

After computing the matrix $S$ and the vector $g$ at time $t$, one can formally define $\dot{\alpha} = -i \, S^{-1} g $ and use any ordinary differential equation (ODE) integrator (see Appendix~\ref{appendix:tdvp_integrators}) to obtain the next set of parameters, at time $t + \delta t$. However, the inverse is often ill defined.

One reason is that Monte Carlo estimates of matrix elements are noisy. Noise accumulates to render the matrix singular by making a small eigenvalues vanish. Therefore, quickly and efficiently obtaining many uncorrelated samples from $ p(\ttheta, t) \propto | \psi _{\alpha (t)} (\ttheta) | ^2$ is crucial. The other reason is that the specific choice of $\psi _\alpha$ introduces redundancy between different parameters, producing linearly dependent or vanishing rows and columns in $S$. Therefore, choosing an efficiently parameterized trial wavefunction is equally important. In practice, adding more parameters to the wavefunction can sometimes unexpectedly reduce accuracy by making $S$ ill conditioned.

In order to move forward with the algorithm, regularization schemes must be used. For ground-state optimization tasks, simply replacing $S \rightarrow S + \epsilon \mathbbm{1}$, for some small positive constant $\epsilon$, often suffices to diminish the effect of small eigenvalues.

However, in this work, we regularize the $S$ matrix by diagonalization $S = U \Sigma U^\dagger$ at each time step. Having obtained eigenvalues $\sigma ^2 _\mu$ such that $\Sigma = \diag (\sigma ^2 _1, \ldots, \sigma ^2 _P)$, we define the pseudoinverse as $ S ^{-1} \approx U \Tilde{\Sigma} ^{-1}  U^\dagger $ with
\begin{equation}
\label{eq:cutoff}
    \Tilde{\Sigma} ^{-1} _{\mu \nu} = \frac{ 1 / \sigma ^2 _\mu} { 1 + \left( \nicefrac{\lambda ^2}{\sigma ^2 _\mu} \right) ^6 } \; \delta _{\mu \nu} \; .
\end{equation}

We heuristically find that the smooth cutoff with a hyperparameter $\lambda ^2$ in Eq.~\ref{eq:cutoff} is superior to traditional pseudoinverses when using adaptive integrators for updating parameters $\alpha $. For more details on regularization, see Appendix~\ref{appendix:regularization}.

After calculating averages in Eq.~\ref{eq:averages} and appropriately regularizing the QGT inverse $S^{-1}$, one can use any external ODE integrator to perform time stepping in the top-level equation $\dot{\alpha} = -i \; S^{-1} g $. In this work, we use the embedded Bogacki-Shampine adaptive solver RK3(2) from the Runge-Kutta family~\cite{Bogacki1989, Butcher2008, Press1992}.

\begin{figure*}[!t]
    \centering
    \includegraphics[width=\textwidth]{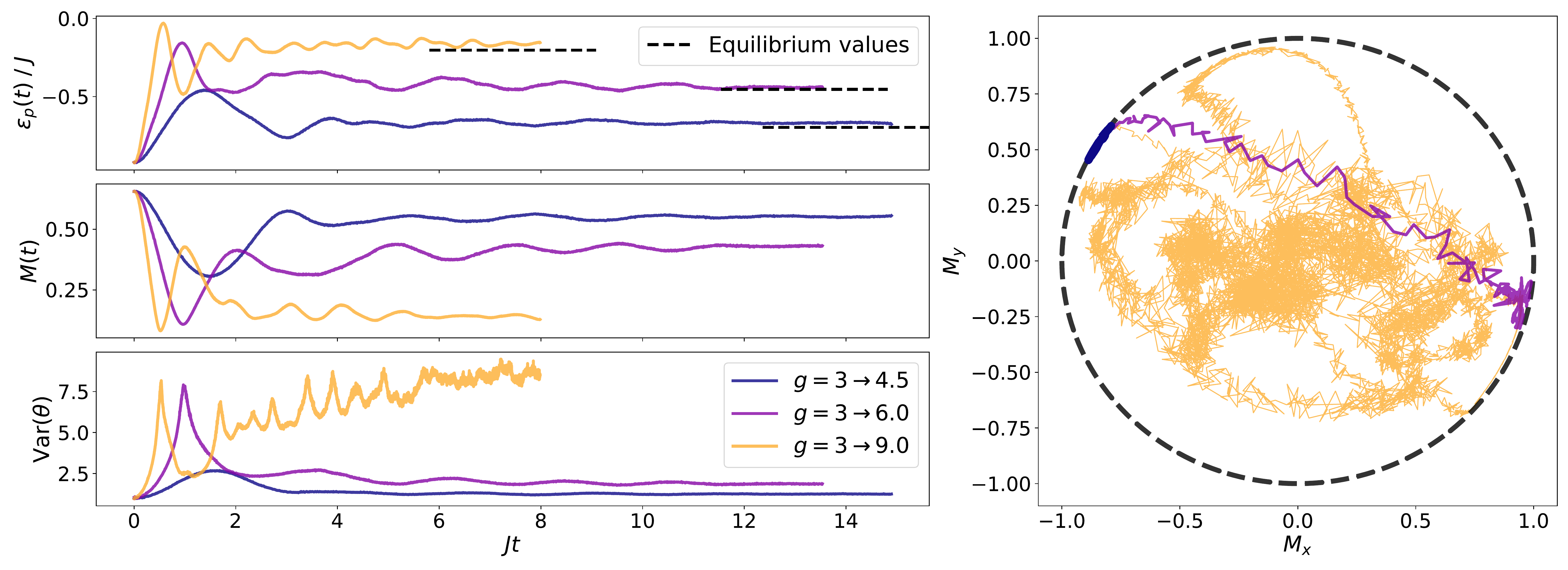}
    \caption{
        Results for different quenches from initial value $g_i = 3$ on a two-dimensional $8 \times 8$ square lattice. 
        \textbf{Left}: Potential energy, magnetization and angular variance as functions of real time. For the small quench to $g_f = 4.5$, we observe the expected behavior with slower approach to the new ordered equilibrium state. Convergence is similar to adiabatic change. The moderate quench to $g_f = 6.0$ exhibits a sharp increase in rotor angle variance is accompanied by a single flip (right panel) in the average magnetization at $t \approx J^{-1}$. For the large quench to $g_f = 9.0$, many rotor flips occur after the first one, indicating much more detailed exploration of the underlying Hilbert space. Convergence to the new equilibrium starts taking place only for $t \gtrsim 5 J^{-1}$.
        \textbf{Right}: A parametric plot of the mean rotor direction. We observe a more thorough exploration of the magnetization sphere for larger quenches.
    }
    \label{fig:observables}
\end{figure*}

\subsection{Hamiltonian Monte Carlo}
\label{sec:hmc}

Hilbert-space averages defined in Eq.~\ref{eq:averages} cannot be evaluated analytically for an arbitrary $\psi _\alpha$. To perform this task in an efficient and scalable way, we employ Hamiltonian Monte Carlo (HMC)~\cite{Neal2012, Betancourt2017} to obtain samples from the distribution $p (\ttheta, t)$ at each time step $t$. We make this choice because HMC offers a systematic way of making large steps in MCMC proposals while still keeping acceptance probabilities high, unlike more conventional approaches like random-walk Metropolis (RWM). This results in a Markov chain with considerably lower autocorrelation times, allowing for treatments of larger systems with less overall runtime spent on sampling.

For a generic probability distribution $p(\ttheta)$, HMC augments the configuration space with artificial momentum variables $\ppi = (\pi _1, \ldots, \pi _N) \sim \mathcal{N}(0, M)$:
\begin{equation}
    \label{eq:hmc_hamiltonian}
    p(\ttheta) \propto \int \dd \ppi \, \exp \left\lbrace -\frac{1}{2} \ppi^\top M^{-1} \ppi + \ln p(\ttheta) \right\rbrace
\end{equation}

\noindent for some choice of a positive-definite \textit{mass matrix} $M$. Interpreting the exponent in Eq.~\ref{eq:hmc_hamiltonian} as an effective classical Hamiltonian $\beta \Tilde{H}(\ttheta, \ppi)$ inducing a Boltzmann weight $e^{-\beta \Tilde{H}}$, Monte Carlo updates can be defined through numerical integration of relevant Hamilton's equations. Owing to insights from statistical physics, we know that a large number of particles in equilibrium following classical equations of motion have precisely this desired Boltzmann distribution.

Given $\ttheta (0)$, $\ppi (0)$ and a small step size $\varepsilon$, a common choice is the leapfrog integrator:
\begin{equation}
\label{eq:leapfrog}
    \begin{gathered}
        \ppi (\tau + \nicefrac{\varepsilon}{2}) = \ppi (\tau) - \frac{\varepsilon}{2} \, \frac{\partial V}{\partial \ttheta } (\ttheta (\tau ) ) \\
        \ttheta (\tau + \varepsilon) = \ttheta(\tau) + \varepsilon \, M^{-1} \, \ppi (\tau + \nicefrac{\varepsilon}{2}) \\
        \ppi (\tau + \varepsilon) = \ppi (\tau + \nicefrac{\varepsilon}{2}) - \frac{\varepsilon}{2} \, \frac{\partial V}{\partial \ttheta } (\ttheta (\tau + \varepsilon) )
    \end{gathered}
\end{equation}

\noindent where $V(\ttheta) = -\ln p(\ttheta)$ and $\tau$ is the fictitious HMC time variable, unrelated to $t$ in Eq.~\ref{eq:tdvp}. This specific integrator is chosen because of its symplectic \cite{Neal2012, Press1992} property -- it conserves energy/probability exactly, allowing for large jumps in the $\ttheta$ space while keeping high acceptance probabilities. We note that higher-order symplectic integrators can be used as well.

After integrating for $L$ steps, the new configuration $(\ttheta (L\varepsilon), \ppi (L\varepsilon))$ is proposed as the next sample in the Markov chain. It is common to apply the Metropolis-Hastings accept-reject step \cite{Metropolis1953, Hastings1970} despite the fact that the new configuration has the same energy (probability) as the initial one. This is done to offset the effects of unwanted numerical errors in the leapfrog scheme, usually improving overall performance for many samples~\cite{Neal2012, Betancourt2017}.

Eqs.~\ref{eq:leapfrog} simulate a swarm of effective classical particles whose positions and momenta follow the desired joint Boltzmann distribution in Eq.~\ref{eq:hmc_hamiltonian}. Discarding all $\ppi$ samples is equivalent to marginalizing the distribution in Eq.~\ref{eq:hmc_hamiltonian}. In practice, randomness is injected by sampling the normal distribution $\ppi (0) \sim \mathcal{N}(0, M)$ each time initial conditions are required for numerical integration.

Choosing the mass matrix $M$, the time step $\varepsilon$ and the integration length $L$ carefully is crucial for efficient exploration of the configuration space. In this work, we chose to set $M$ and $\varepsilon$ automatically, by using heuristically proven \cite{Nesterov2009, Hoffman2011, Betancourt2017} algorithms operating samples from an extended warmup phase for each Markov chain individually. Integration length $L$ was treated as a hyperparameter. For more details and specific values, see Appendix~\ref{appendix:hmc}.

\begin{figure*}[!t]
    \centering
    \begin{minipage}{0.49\textwidth}
        \includegraphics[width=\linewidth]{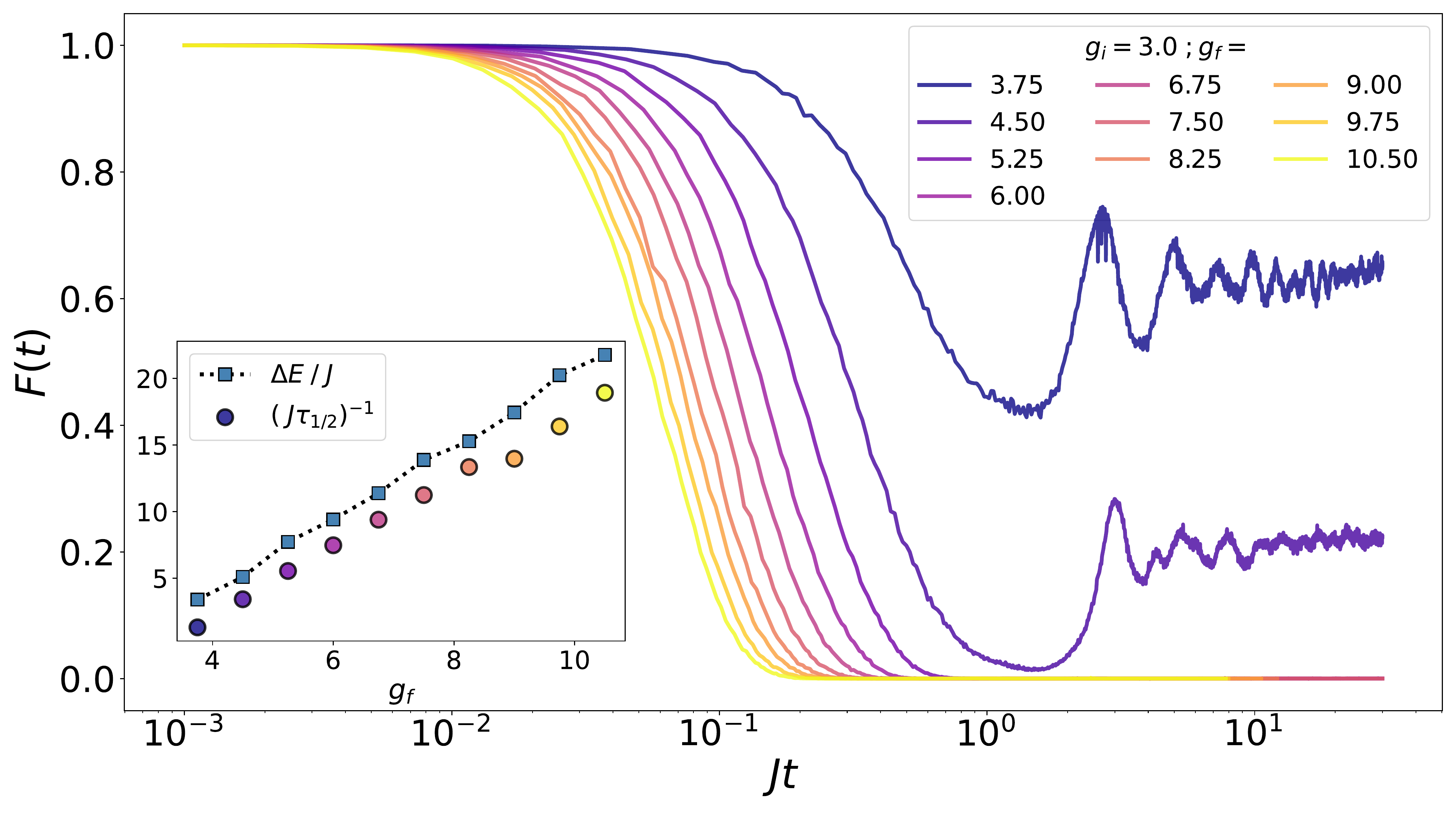}
    \end{minipage}\quad
    \begin{minipage}{0.49\textwidth}
        \includegraphics[width=\linewidth]{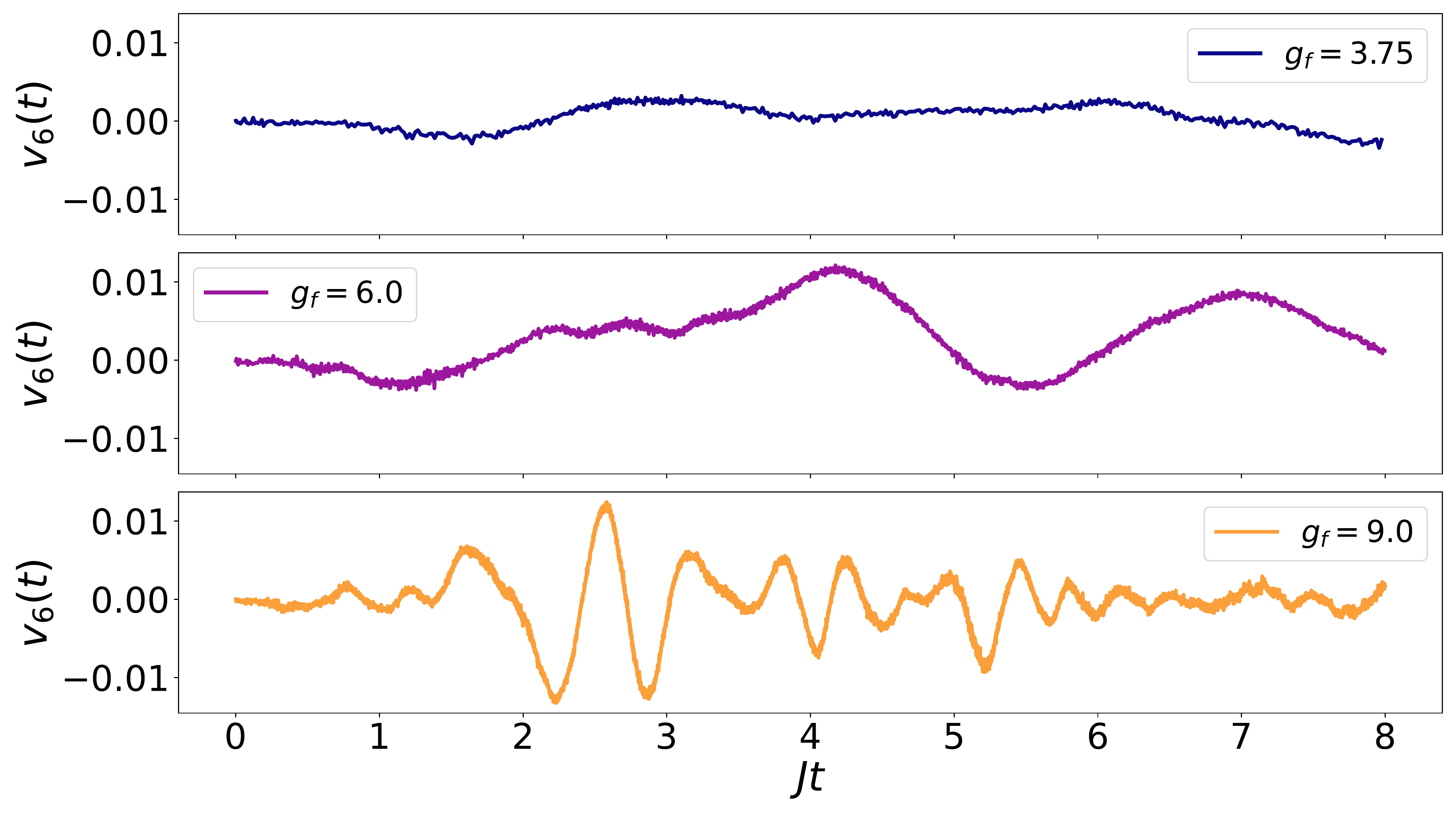}
    \end{minipage}
    \caption{
        Fidelity and vorticity as functions of time.
        \textbf{Left}: Time-dependent many-body fidelity $F(t)$ defined in Eq.~\ref{eq:fidelity}, for a number of quenches. For trajectories quenching to values of $g_f$ in the same equilibrium phase, we see convergence to nonzero values at late times. Conversely, trajectories with $g_f > g_c$ converge to $F(t \rightarrow \infty ) = 0$. Additionally, $\tau_{\nicefrac{1}{2}}$ (the time it takes for fidelity to decrease by $50\%$) is shown to scale linearly with $g$ in agreement with the appropriate uncertainty relation $\Delta E \Delta t \geq \nicefrac{1}{2}$.
        \textbf{Right}: The onset of vorticity (defined in Eq.~\ref{eq:vorticity}) for three quenches of increasing magnitude.
    }
    \label{fig:fidelities_vorticities}
\end{figure*}

\subsection{The trial wavefunction}

In this work, we use a variant of the standard convolutional neural network (CNN) architecture \cite{LeCun2015, Carleo2019} to model $\psi _\alpha (\ttheta)$. Our approach is built on those of Refs.~\cite{Schmitt2020, Pescia2022}. Specifically, we set
\begin{equation}
\label{eq:cnn_1}
    \ln \psi _\alpha (\ttheta) = \frac{1}{\sqrt{2 K N}} \sum _{c=1} ^{2 K} \sum _k \left[ w ^c _D \ast h ^c _{D-1} (\ttheta ) \right] _k \; ,
\end{equation}

\noindent where $\ast$ denotes a convolution over lattice indices $k$ and $c = 1, \ldots, 2 K$ is the channel index. Features $h ^c _{D-1} (\ttheta )$ are the output of $D-1$-layer CNN defined by:
\begin{equation}
    h ^c _d (\ttheta ) = f _d \left( b^c _d + \sum _{c' = 1} ^{2 K} w^{c c'} _d \ast  h ^{c'} _{d-1} (\ttheta ) \right)
\end{equation}

\noindent with an elementwise nonlinear activation function $f _d$, biases $b_d$, and weights $w_d$ at layer $d$. We include all weights and biases into the set of trainable parameters $\alpha $ and use automatic differentiation (AD) techniques to obtain all derivatives $\OO _\mu$ required for evaluation of Eqs.~\ref{eq:averages}. For CNN inputs $h_0$, we concatenate the following features:
\begin{equation}
    h_0 = \left\{ (\cos n \theta _k, \sin n \theta _k) \; \Big{|} \; n = 1, \ldots, K \right\}
\end{equation}

\noindent along the channel axis, as illustrated on Fig.~\ref{fig:diagram}. This construction allows us to include a limited number of higher Fourier modes \textit{a priori}, improving ansatz expressivity in a controlled way. In this work, we set $D=2$, $K=4$ for larger two-dimensional ($8 \times 8$) experiments and $K=1$ for smaller systems.

To maintain analytic dependence on parameters $\alpha$, we restrict the CNN nonlinearities $f_d$ to polynomial functions. The Taylor expansion of the logarithm of the zeroth-order modified Bessel function of the first kind is used:
\begin{equation}
\label{eq:taylor_activation}
    \ln I _0 (z) = \frac{z^2}{4} - \frac{z^4}{64} + \frac{z^6}{576} + \mathcal{O}(z^8) \; .
\end{equation}

\noindent This particular activation function choice is motivated by the appearance of $I_0$ in the version of the restricted Boltzmann machine (RBM) adapted to the QRM in Ref.~\cite{Stokes2021}. This approach has the advantage of maintaining the holomorphic dependence of $\psi _\alpha$ on $\alpha$ and preserving the form of Eqs.~\ref{eq:averages}.

In this work, we focus on a simple two-layer CNN ansatz to control the number of parameters $P$. In addition nontrivially affecting the QGT inverse (see subsection~\ref{sec:variational_sim}), the cost to diagonalize the QGT in order to regularize the inverse in Eq.~\ref{eq:cutoff} grows as $\mathcal{O}(P^3)$. Heuristically, we also find that introducing more parameters $\alpha$ requires more Monte Carlo samples to correctly resolve the relevant averages in Eq.~\ref{eq:averages} and does not significantly contribute to simulation accuracy in our case. A systematic investigation of larger neural-network architecture details is left for future work.

\begin{figure*}[!t]
    \centering
    \includegraphics[width=\linewidth]{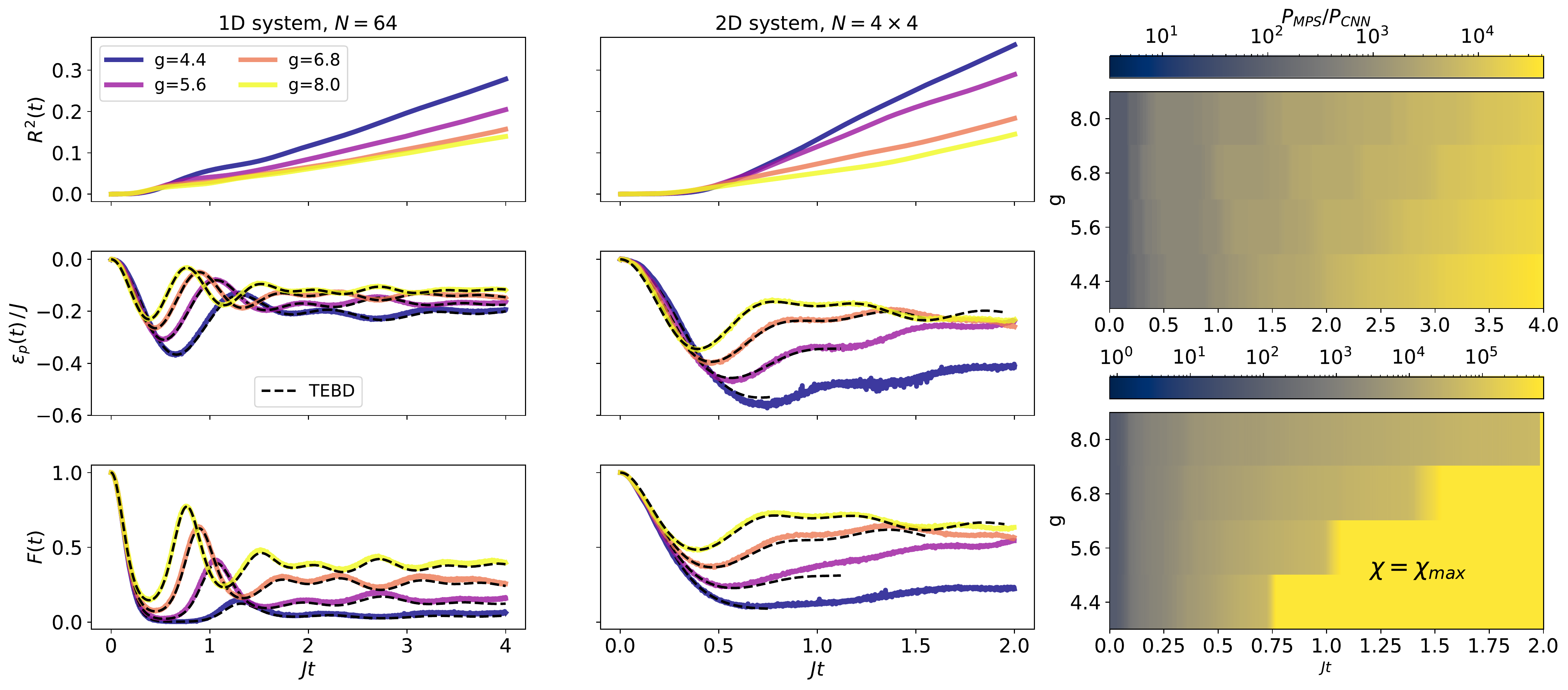}
    \caption{
        One- and two-dimensional benchmarks and comparison with tensor-network data. Evolution was performed starting from a coherent superposition state $\ket{\psi (0)} \propto \int \dd \ttheta \ket{\ttheta} $. Results are compared with the TEBD tensor-network algorithm evolving a matrix product state (MPS) in the conjugate angular-momentum eigenbasis (see Sec.~\ref{sec:comparison} and Appendix~\ref{appendix:tensors}).
        \textbf{Left}:
            A one-dimensional benchmark on a chain with $N=64$ rotors and open boundary conditions.
        \textbf{Center}:
            A two-dimensional benchmark of the t-VMC method on a $4 \times 4$ lattice and open boundary conditions. We note that disagreement between t-VMC and TEBD results appears as the maximum bond dimension $\chi _\text{max}$ is reached. Singular value cutoff of $10^{-12}$ was used.
        \textbf{Right}:
            The growing number of MPS parameters $P_\text{MPS}$ associated with the increasing bond dimension $\chi$ is plotted in units of the number of the CNN parameter count $P_\text{CNN}$ as a function of time. One- and two-dimensional cases are compared. A cutoff of $\chi _\text{max} = 1000$ was reached in the 2D system for the singular value cutoff of $10^{-9}$.
    }
    \label{fig:methods}
\end{figure*}

\section{Results}
\label{sec:results}

In this section, we study dynamical properties of several observables of the QRM, focusing on the two-dimensional model. A series of benchmarks in one and two dimensions can be found in section~\ref{sec:comparison}.

We simulate the effects instantaneous \textit{quenches} of the coupling constant $g$ in Eq.~\ref{eq:hamiltonian}. Specifically, we initialize parameters $\alpha$ of the ansatz $\psi _\alpha$ illustrated on Fig.~\ref{fig:diagram} to the ground state of the QRM Hamiltonian with $g=g_i$ using imaginary-time variational Monte Carlo (VMC)~\cite{Carleo2019, Becca2017} methods. We then simulate real-time dynamics under $g=g_f$. In this work, we focus on quenches from the ordered phase to the disordered: $g_i < g_c < g_f$.

In Fig.~\ref{fig:observables}, we choose a square $8 \times 8$ lattice, tracking the dynamics of the potential energy density
\begin{equation}
\label{eq:pe_density}
    \epsilon _\text{p} (t) = - \frac{J}{N} \avg{ \sum_{\langle k, l \rangle} \hat{\vb{n}} _k \cdot \hat{\vb{n}} _l } _t
\end{equation}

\noindent and the average magnetization magnitude $M$
\begin{equation}
\label{eq:magnetization}
    M(t) = \frac{1}{N} \avg{ \left\vert \sum\nolimits_k \hat{\vb{n}} _k \right\vert } _t \; ,
\end{equation}

\noindent along with its $x$, $y$ components defined by $\vb{M} = N^{-1} \sum _k \avg{\hat{\vb{n}} _k} _t$. Averages $\avg{\cdot}_t$ are performed with respect to the ansatz state at time $t$. In addition, corresponding circular variances were defined as $\Var(\theta _k) = -2 \ln \left| \avg{\hat{\vb{n}} _k} _t \right|$ and averaged over the lattice index $k$.

These observables were chosen as a proxy for thermalization. Across a wide range of quenches we observe convergence to their respective equilibrium values at $g=g_f$, see Fig.~\ref{fig:observables}. We observe two distinct dynamical regimes in relation to the quantum critical point $g_c \approx 4.25$, when $g_i < g_c$. For small quenches (left column of Fig.~\ref{fig:observables}) we see the expected outcome -- slower equilibriation with only small fluctuations in the direction of the magnetization. However, for moderate to large quenches in Fig.~\ref{fig:observables}, we observe a (transient) demagnetization of the sample and convergence to a new equilibrium state. 

In addition, we define a measure of average vorticity
\begin{equation}
\label{eq:vorticity}
    v(A) = \frac{1}{|A|} \int _A \dd \vb{a} \cdot \nabla \times \hat{\vb{n}} = \frac{1}{|A|} \oint _{\partial A} \dd \bm{\ell} \cdot \hat{\vb{n}}  
\end{equation}

\noindent over a surface $A$ with edge $\partial A$ on the lattice. Using Stokes' theorem, we rewrite the expression as a contour integral over $\partial A$ in the positive direction. On Fig.~\ref{fig:fidelities_vorticities} (right panel), we plot $v(A)$ averaged over all $n_\ell$ square $\ell \times \ell$ surfaces: $v_\ell = n_\ell ^{-1} \sum _{\left| A \right| = \ell ^2} v(A)$. As expected, we find almost zero vorticity for quenches in the ordered phase, while larger fluctuations are generated for quenches across the critical point. We postpone a detailed analysis for future work.

Aside from local observables, such as energy and magnetization, one also has access to global observables such as the Loschmidt echo. The latter has some interesting properties in the context of dynamical phase transitions~\cite{heyl13} and quantum chaos~\cite{schmittsels19}. The Loschmidt echo expresses the quantum state overlap between the initial state and some time-evolved state. In general, the fidelity $F(\Psi, \Phi)$ between two generic normalized quantum states $\Psi$ and $\Phi$ is defined as $F(\Psi, \Phi) = \left\vert \braket{\Psi}{\Phi} \right\vert ^2$. For real-time evolution, we expect the fidelity $F(\Psi(t=0), \Psi(t))$ to decay as a function of time $t$, for any given initial state $\ket{\Psi(t=0)}$.

To evaluate this quantity using Monte Carlo sampling of unnormalized ansatz wavefunctions $ \psi (\ttheta, t) =  \psi _{\alpha (t)} (\ttheta)$, we rewrite the fidelity definition as:
\begin{equation}
\label{eq:fidelity}
    F(t) = \avg{\frac{\psi (\ttheta , t)}{\psi (\ttheta, 0)}} _{\ttheta \sim |\psi (\cdot, 0)| ^2} \avg{\frac{\psi (\ttheta, 0)}{\psi (\ttheta, t)}} _{\ttheta \sim |\psi (\cdot, t)| ^2} \; ,
\end{equation}

\noindent following Refs.~\cite{Medvidovic2021, Jonsson2018}. The expression in Eq.~\ref{eq:fidelity} is manifestly independent of the normalization factor. In practice, we take the real part of Eq.~\ref{eq:fidelity} to discard the small nonzero imaginary part coming from finite-sample estimates of the two factors. In addition, we calculate and store both factors in log space to preserve accuracy and maintain numerical stability.

As expected, we find that that the return probability (or fidelity in short) decays quickly with time, as illustrated in Fig.~\ref{fig:fidelities_vorticities} (left panel). For smaller quenches, the fidelity shoots back up to a nonzero value suggesting a finite overlap between the initial state the long time "equilibrium" state after the quench. The latter may be interpreted as a signature of quenching between two Hamiltonians in the ordered phase. 

As a measure of the fidelity decay, we introduce another time scale $\tau _{\nicefrac{1}{2}}$ defined as the time needed for the fidelity to decrease by $50\%$. We observe that $\tau _{\nicefrac{1}{2}}$ has increased linearly with the quench $g_f$. This result matches basic estimates given by the second-order short-time expansion of $F(t)$ and uncertainty relation $\Delta E \Delta t \geq \nicefrac{1}{2} $. Therefore, fidelity decay time can be lower bounded by $\Delta E ^{-1}$, estimated using samples from the initial state $\psi _{\alpha(0)}$~\cite{Mandelstam1991}. Reference points from this calculation are presented in Fig.~\ref{fig:fidelities_vorticities} (left, inset). This comparison demonstrates that the t-VMC method can be used to estimate quantities of experimental interest for system sizes unreachable by other wavefunction-based methods.

\subsection{Benchmarks}
\label{sec:comparison}

To substantiate our results, we perform a series of benchmarks and compare results to tensor-network simulations for a one- and two-dimensional versions of the model. In particular we benchmark the results with the time-evolving block decimation (TEBD)~\cite{Vidal2003, Vidal2004} algorithm. For all benchmarks, states were initialized to the coherent superposition of all basis states $\ket{\psi (0)} \propto \int \dd \ttheta \ket{\ttheta}$ by explicitly setting the final convolution kernel $w ^c _D$ (Eq.~\ref{eq:cnn_1}) to zero. All presented tensor-network simulations have been performed with a fixed singular value cutoff. Convergence within the matrix product state (MPS) variational manifold has been confirmed by repeating simulations with larger cutoff values.

We organize numerical benchmarks as follows. First, we compare t-VMC results with TEBD for an extended one-dimensional and a smaller two-dimensional system. Practical error estimates are defined. Then, we turn to examining effects of key hyperparameters in the t-VMC approach and show evidence of self-consistent convergence.

Following Refs.~\cite{Schmitt2020, Carleo2017}, we use the following figure of merit:
\begin{equation}
\label{eq:full_residual}
    r (t) = \frac{\mathcal{D} \left( \psi (t +\delta t), e^{-i H \delta t} \, \psi (t) \right)}{\mathcal{D} \left( \psi (t), e^{-i H \delta t} \, \psi (t) \right)}
\end{equation}

\noindent where $ \ket{\psi (t)} = \ket{\psi _{\alpha (t)}}$. In Eq.~\ref{eq:full_residual}, $\mathcal{D}(\cdot, \cdot)$ represents the Fubini-Study distance on the Hilbert space $\mathcal{H}$. We estimate $r^2(t)$ at each time $t$ using HMC samples from the ansatz (see Ref.~\cite{Schmitt2020} and Appendix~\ref{appendix:performance}). Intuitively, $r^2 (t)$ measures an appropriately normalized measure of deviation between the full state $e^{-i H \delta t} \, \ket{\psi (t)} $ after one time step $\delta t$ and its projection onto the variational manifold $\ket{\psi _{\alpha (t + \delta t)}}$. We plot the integrated error
\begin{equation}
\label{eq:R2}
    R^2 (t) = \int _0 ^t r^2(s) \, \dd s
\end{equation}

\noindent to reflect error propagation through time as accurately as possible. We remark that the integrated-squared error in Eq.~\ref{eq:R2} should be interpreted an upper bound on the square of the integrated error $R (t) = \int _0 ^t r(s) \, \dd s$ due to the triangle inequality.

In Fig.~\ref{fig:methods} (left), we show that this algorithm performs well on a one-dimensional system of $N=64$ rotors where the growth of the so-called \textit{bond dimension} $\chi$ is limited. Convergence to appropriate equilibrium values is reached for both methods with good agreement at intermediate times for the dynamics of potential energy density $\epsilon _\text{p} (t)$ and the Loschmidt echo $F(t)$. The \textit{integrated residual} $R^2 (t)$ grows more rapidly for lower values of $g$. This is expected because the initial state $\psi (0)$ has lower energy for larger values of $g$ in the QRM Hamiltonian, Eq.~\ref{eq:hamiltonian}, representing a more typical state in the disordered phase.

In contrast to the one-dimensional (1D) case, in Fig.~\ref{fig:methods} (center), we observe that the TEBD method exponentially grows the MPS bond dimension $\chi$ past the cutoff $\chi _\text{max} = 1000$ at relatively short times. We plot the number of parameters $P_\text{MPS}$ in the MPS as a function of time in the right panel of Fig.~\ref{fig:methods}, in units of the number of parameters $P_\text{CNN}$ in the CNN ansatz presented in this work. We see qualitative agreement between the two methods for early times, before $\chi$ grows to the point where further simulation is numerically prohibitively expensive.

\begin{figure}[!t]
    \centering
    \includegraphics[width=\linewidth]{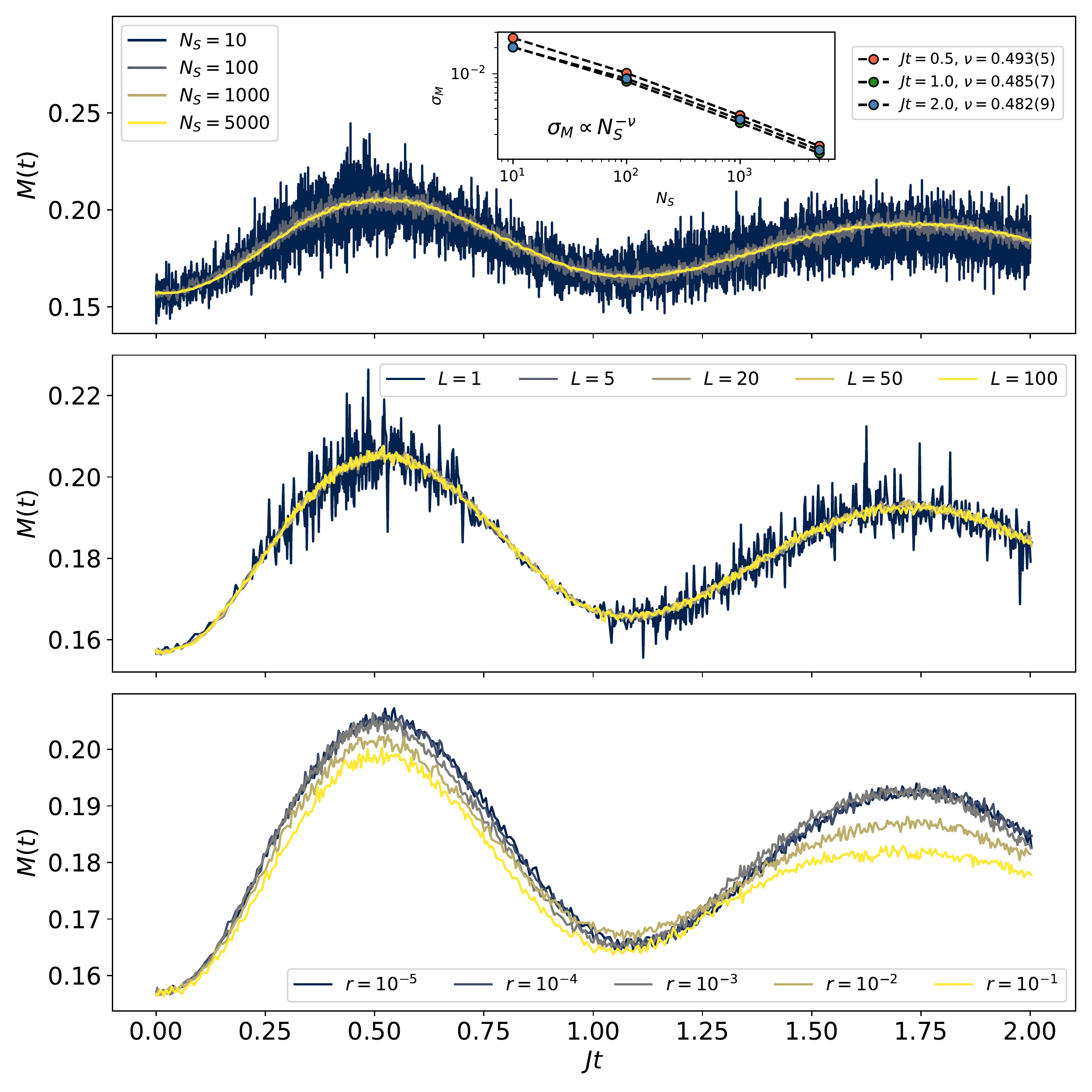}
    \caption{
        Effects of key hyperparameters on magnetization measurements. All experiments were performed on a one-dimensional chain with $N=32$.
        \textbf{Top}:
            Effects on magnetization estimates by varying the number of HMC samples $N_s$. Errors were estimated using bootstrap resampling independently at different times show expected scaling $\sigma _M \propto N _{\text{s}} ^{-\nicefrac{1}{2}}$ in all cases.
        \textbf{Middle}:
            Variance change in magnetization estimates by varying the number of leapfrog integrator steps $L$ between HMC proposals. In the $L \rightarrow 1$ limit, HMC approaches the random-walk Metropolis sampler.
        \textbf{Bottom}:
            Bias increase associated with changing the $\lambda ^2$ cutoff parameter in Eq.~\ref{eq:cutoff}.
    }
    \label{fig:hyperparams}
\end{figure}

In Fig.~\ref{fig:hyperparams} we show evidence that the variance of observables is controllable through the most important Monte Carlo (HMC) hyperparameters while the bias is mostly controlled by different regularizations of the $S$-matrix inverse (Eq.~\ref{eq:averages}). In the top panel of Fig.~\ref{fig:hyperparams}, we see that the standard deviation of the estimator for total magnetization $M(t)$ scales with the number of HMC samples $N_\text{s}$ in an expected way: $\sigma _M \propto N_\text{s} ^{-\nicefrac{1}{2}}$ for three different times during the evolution.

In addition, we report that heuristically varying the number of leapfrog integration steps $L$ increases estimator variances the most around segments of trajectories with higher curvature, as evidenced by the middle panel of Fig.~\ref{fig:hyperparams}. Intuitively, in the limit of $L \rightarrow 1$ and small leapfrog step sizes $\varepsilon$, HMC approaches random-walk Metropolis sampling (see Ref.~\cite{Betancourt2017} and Appendix~\ref{appendix:hmc}) which suffers from lower acceptance rates and longer mixing times in cases of sharply peaked target distributions. We observe that even a moderate increase to $L \approx 10$ accompanied by automatic hyperparameter tuning described in Sec.~\ref{sec:hmc} considerably reduces variance.

Finally, we explore the effects of $S$-matrix regularization (Eq.~\ref{eq:cutoff}). In practice, we set $\lambda ^2$ itself in an adaptive manner each iteration:
\begin{equation}
    \lambda ^2 = \lambda ^2 (\sigma ^2 _1, \ldots, \sigma ^2 _P ) = \max \left(a_c, r_c \times \max _\mu (\sigma ^2 _\mu) \right)
\end{equation}

\noindent depending on the $S$-matrix spectrum. In the bottom panel of Fig.~\ref{fig:hyperparams}, we see that, for a fixed $a_c = 10^{-5}$, increasing $r_c$ leads to increasing the estimator bias. Excluding relevant eigenvalues from participating in time evolution through Eq.~\ref{eq:cutoff} can lead to a failure to capture parts of relevant physics.

Overall, both t-VMC and TEBD algorithms predict similar dynamical behavior of the potential energy density (Eq.~\ref{eq:pe_density}) and the fidelity (Eq.~\ref{eq:fidelity}), as shown on Fig.~\ref{fig:methods}. However, the number of parameters in the MPS grows exponentially due to entropy buildup during time evolution. Tensor-network real-time evolution algorithms \cite{Haegeman2011, Haegeman2016} based on MPS or two-dimensional architectures such as projected entangled pair states (PEPS) \cite{Verstraete2004, Czarnik2019} face several challenges to extend to late times and higher dimensions. Incorporating continuous degrees of freedom exacerbates the problem -- tensor network algorithms are limited to using the locally truncated eigenbasis of the angular momentum operator $L_k$ in the QRM Hamiltonian in Eq.~\ref{eq:hamiltonian}, in contrast to the t-VMC method (see Appendix~\ref{appendix:tensors}).

\section{Conclusion}
\label{sec:conclusion}

We present a method to approximate unitary dynamics of continuous-variable quantum many-body systems, based on custom neural-network quantum states. The approach employs Hamiltonian Monte Carlo sampling and custom regularization of the quantum geometric tensor. The method was benchmarked on quench dynamics of two-dimensional quantum rotors. We indicated that our calculations are able to access nonlocal quantities like the return probability. Good agreement was found with tensor-network-based TEBD simulations for the case of one-dimensional systems of comparable size. Finally, we showed evidence that the method is controlled by a handful of key hyperparameters. Our approach paves the way for accurate nonequilibrium simulations of continuous systems at previously unexplored system sizes and evolution times, bridging the gap between simulation and experiment. 

\section{Acknowledgements}
\label{sec:acknowledgements}

M.~M. acknowledges insightful discussions with Filippo Vicentini about t-VMC regularization, Bob Carpenter about the role of circular geometry in Monte Carlo sampling and Hamiltonian Monte Carlo details. In addition, discussions with Sandro Sorella about the infinite variance problem and James Stokes about different ansatze were very helpful for fine tuning simulations. MM also acknowledges support from the CCQ graduate fellowship in computational quantum physics. The Flatiron Institute is a division of the Simons Foundation. D.~S. was supported by AFOSR: Grant No. FA9550-21-1-0236 and NSF: Grant No. OAC-2118310.

\subsection*{Software libraries}

The code used in this work has been packaged into an installable library and is publicly available to reproduce any results in this work or explore new ones: \href{https://github.com/Matematija/continuous-vmc}{github.com/Matematija/continuous-vmc}.

It was built on JAX~\cite{jax2018github} for array manipulations, automatic differentiation for sampling and optimization and GPU support, Flax~\cite{flax2020github} for neural-network construction and manipulation and NumPy~\cite{numpy} and SciPy~\cite{scipy} for CPU array manipulations. Matplotlib~\cite{matplotlib} was used to produce figures.

\bibliographystyle{naturemag}
\bibliography{references}

\begin{thebibliography}{10}
\expandafter\ifx\csname url\endcsname\relax
  \def\url#1{\texttt{#1}}\fi
\expandafter\ifx\csname urlprefix\endcsname\relax\def\urlprefix{URL }\fi
\providecommand{\bibinfo}[2]{#2}
\providecommand{\eprint}[2][]{\url{#2}}

\bibitem{rabitz93}
\bibinfo{author}{Warren, W.~S.}, \bibinfo{author}{Rabitz, H.} \&
  \bibinfo{author}{Dahleh, M.}
\newblock \bibinfo{title}{Coherent control of quantum dynamics: The dream is
  alive}.
\newblock \emph{\bibinfo{journal}{Science}} \textbf{\bibinfo{volume}{259}},
  \bibinfo{pages}{1581--1589} (\bibinfo{year}{1993}).

\bibitem{rmpPolkovnikov11}
\bibinfo{author}{Polkovnikov, A.}, \bibinfo{author}{Sengupta, K.},
  \bibinfo{author}{Silva, A.} \& \bibinfo{author}{Vengalattore, M.}
\newblock \bibinfo{title}{Colloquium: Nonequilibrium dynamics of closed
  interacting quantum systems}.
\newblock \emph{\bibinfo{journal}{Rev. Mod. Phys.}}
  \textbf{\bibinfo{volume}{83}}, \bibinfo{pages}{863--883}
  (\bibinfo{year}{2011}).
\newblock \urlprefix\url{https://link.aps.org/doi/10.1103/RevModPhys.83.863}.

\bibitem{cavalleri21}
\bibinfo{author}{Budden, M.} \emph{et~al.}
\newblock \bibinfo{title}{Evidence for metastable photo-induced
  superconductivity in k3c60}.
\newblock \emph{\bibinfo{journal}{Nature Physics}}
  \textbf{\bibinfo{volume}{17}}, \bibinfo{pages}{611--618}
  (\bibinfo{year}{2021}).
\newblock \urlprefix\url{https://doi.org/10.1038/s41567-020-01148-1}.

\bibitem{Ebbesen19}
\bibinfo{author}{Thomas, A.} \emph{et~al.}
\newblock \bibinfo{title}{Tilting a ground-state reactivity landscape by
  vibrational strong coupling}.
\newblock \emph{\bibinfo{journal}{Science}} \textbf{\bibinfo{volume}{363}},
  \bibinfo{pages}{615--619} (\bibinfo{year}{2019}).
\newblock
  \urlprefix\url{https://www.science.org/doi/abs/10.1126/science.aau7742}.
\newblock \eprint{https://www.science.org/doi/pdf/10.1126/science.aau7742}.

\bibitem{Zhang17}
\bibinfo{author}{Zhang, J.} \emph{et~al.}
\newblock \bibinfo{title}{Observation of a many-body dynamical phase transition
  with a 53-qubit quantum simulator}.
\newblock \emph{\bibinfo{journal}{Nature}} \textbf{\bibinfo{volume}{551}},
  \bibinfo{pages}{601--604} (\bibinfo{year}{2017}).
\newblock \urlprefix\url{https://doi.org/10.1038/nature24654}.

\bibitem{Sivak2022}
\bibinfo{author}{Sivak, V.~V.} \emph{et~al.}
\newblock \bibinfo{title}{Model-free quantum control with reinforcement
  learning}.
\newblock \emph{\bibinfo{journal}{Physical Review X}}
  \textbf{\bibinfo{volume}{12}}, \bibinfo{pages}{011059}
  (\bibinfo{year}{2022}).
\newblock \urlprefix\url{https://link.aps.org/doi/10.1103/PhysRevX.12.011059}.

\bibitem{Porotti2022}
\bibinfo{author}{Porotti, R.}, \bibinfo{author}{Essig, A.},
  \bibinfo{author}{Huard, B.} \& \bibinfo{author}{Marquardt, F.}
\newblock \bibinfo{title}{Deep reinforcement learning for quantum state
  preparation with weak nonlinear measurements}.
\newblock \emph{\bibinfo{journal}{Quantum}} \textbf{\bibinfo{volume}{6}},
  \bibinfo{pages}{747} (\bibinfo{year}{2022}).

\bibitem{Metz2022}
\bibinfo{author}{Metz, F.} \& \bibinfo{author}{Bukov, M.}
\newblock \bibinfo{title}{Self-correcting quantum many-body control using
  reinforcement learning with tensor networks}  (\bibinfo{year}{2022}).
\newblock \urlprefix\url{http://arxiv.org/abs/2201.11790}.

\bibitem{Bukov2018}
\bibinfo{author}{Bukov, M.} \emph{et~al.}
\newblock \bibinfo{title}{Reinforcement learning in different phases of quantum
  control}.
\newblock \emph{\bibinfo{journal}{Physical Review X}}
  \textbf{\bibinfo{volume}{8}}, \bibinfo{pages}{031086} (\bibinfo{year}{2018}).
\newblock \urlprefix\url{https://link.aps.org/doi/10.1103/PhysRevX.8.031086}.

\bibitem{zaletel15}
\bibinfo{author}{Zaletel, M.~P.}, \bibinfo{author}{Mong, R. S.~K.},
  \bibinfo{author}{Karrasch, C.}, \bibinfo{author}{Moore, J.~E.} \&
  \bibinfo{author}{Pollmann, F.}
\newblock \bibinfo{title}{Time-evolving a matrix product state with long-ranged
  interactions}.
\newblock \emph{\bibinfo{journal}{Phys. Rev. B}} \textbf{\bibinfo{volume}{91}},
  \bibinfo{pages}{165112} (\bibinfo{year}{2015}).
\newblock \urlprefix\url{https://link.aps.org/doi/10.1103/PhysRevB.91.165112}.

\bibitem{wurtz18}
\bibinfo{author}{Wurtz, J.}, \bibinfo{author}{Polkovnikov, A.} \&
  \bibinfo{author}{Sels, D.}
\newblock \bibinfo{title}{Cluster truncated wigner approximation in strongly
  interacting systems}.
\newblock \emph{\bibinfo{journal}{Annals of Physics}}
  \textbf{\bibinfo{volume}{395}}, \bibinfo{pages}{341--365}
  (\bibinfo{year}{2018}).
\newblock
  \urlprefix\url{https://www.sciencedirect.com/science/article/pii/S0003491618301647}.

\bibitem{Czarnik2019}
\bibinfo{author}{Czarnik, P.}, \bibinfo{author}{Dziarmaga, J.} \&
  \bibinfo{author}{Corboz, P.}
\newblock \bibinfo{title}{Time evolution of an infinite projected entangled
  pair state: An efficient algorithm}.
\newblock \emph{\bibinfo{journal}{Physical Review B}}
  \textbf{\bibinfo{volume}{99}}, \bibinfo{pages}{035115}
  (\bibinfo{year}{2019}).
\newblock \urlprefix\url{https://link.aps.org/doi/10.1103/PhysRevB.99.035115}.

\bibitem{hubig20}
\bibinfo{author}{Hubig, C.}, \bibinfo{author}{Bohrdt, A.},
  \bibinfo{author}{Knap, M.}, \bibinfo{author}{Grusdt, F.} \&
  \bibinfo{author}{Cirac, I.}
\newblock \bibinfo{title}{Evaluation of time-dependent correlators after a
  local quench in ipeps: hole motion in the t-j model}.
\newblock \emph{\bibinfo{journal}{SciPost Physics}}
  \textbf{\bibinfo{volume}{8}}, \bibinfo{pages}{021} (\bibinfo{year}{2020}).
\newblock \urlprefix\url{https://scipost.org/10.21468/SciPostPhys.8.2.021}.

\bibitem{zou20}
\bibinfo{author}{Zhou, Y.}, \bibinfo{author}{Stoudenmire, E.~M.} \&
  \bibinfo{author}{Waintal, X.}
\newblock \bibinfo{title}{What limits the simulation of quantum computers?}
\newblock \emph{\bibinfo{journal}{Phys. Rev. X}} \textbf{\bibinfo{volume}{10}},
  \bibinfo{pages}{041038} (\bibinfo{year}{2020}).
\newblock \urlprefix\url{https://link.aps.org/doi/10.1103/PhysRevX.10.041038}.

\bibitem{Carleo2012}
\bibinfo{author}{Carleo, G.}, \bibinfo{author}{Becca, F.},
  \bibinfo{author}{Schiró, M.} \& \bibinfo{author}{Fabrizio, M.}
\newblock \bibinfo{title}{Localization and glassy dynamics of many-body quantum
  systems}.
\newblock \emph{\bibinfo{journal}{Scientific Reports}}
  \textbf{\bibinfo{volume}{2}}, \bibinfo{pages}{243} (\bibinfo{year}{2012}).
\newblock \urlprefix\url{http://www.nature.com/articles/srep00243}.

\bibitem{Carleo2017_2}
\bibinfo{author}{Carleo, G.}, \bibinfo{author}{Cevolani, L.},
  \bibinfo{author}{Sanchez-Palencia, L.} \& \bibinfo{author}{Holzmann, M.}
\newblock \bibinfo{title}{Unitary dynamics of strongly interacting bose gases
  with the time-dependent variational monte carlo method in continuous space}.
\newblock \emph{\bibinfo{journal}{Physical Review X}}
  \textbf{\bibinfo{volume}{7}}, \bibinfo{pages}{031026} (\bibinfo{year}{2017}).
\newblock
  \urlprefix\url{https://journals.aps.org/prx/abstract/10.1103/PhysRevX.7.031026}.

\bibitem{Schmitt2020}
\bibinfo{author}{Schmitt, M.} \& \bibinfo{author}{Heyl, M.}
\newblock \bibinfo{title}{Quantum many-body dynamics in two dimensions with
  artificial neural networks}.
\newblock \emph{\bibinfo{journal}{Physical Review Letters}}
  \textbf{\bibinfo{volume}{125}}, \bibinfo{pages}{100503}
  (\bibinfo{year}{2020}).
\newblock
  \urlprefix\url{https://link.aps.org/doi/10.1103/PhysRevLett.125.100503}.

\bibitem{Hofmann2022}
\bibinfo{author}{Hofmann, D.}, \bibinfo{author}{Fabiani, G.},
  \bibinfo{author}{Mentink, J.}, \bibinfo{author}{Carleo, G.} \&
  \bibinfo{author}{Sentef, M.}
\newblock \bibinfo{title}{Role of stochastic noise and generalization error in
  the time propagation of neural-network quantum states}.
\newblock \emph{\bibinfo{journal}{SciPost Physics}}
  \textbf{\bibinfo{volume}{12}}, \bibinfo{pages}{165} (\bibinfo{year}{2022}).
\newblock \urlprefix\url{https://scipost.org/10.21468/SciPostPhys.12.5.165}.

\bibitem{Barison2021}
\bibinfo{author}{Barison, S.}, \bibinfo{author}{Vicentini, F.} \&
  \bibinfo{author}{Carleo, G.}
\newblock \bibinfo{title}{An efficient quantum algorithm for the time evolution
  of parameterized circuits}.
\newblock \emph{\bibinfo{journal}{Quantum}} \textbf{\bibinfo{volume}{5}},
  \bibinfo{pages}{512} (\bibinfo{year}{2021}).

\bibitem{Czischek18}
\bibinfo{author}{Czischek, S.}, \bibinfo{author}{G\"arttner, M.} \&
  \bibinfo{author}{Gasenzer, T.}
\newblock \bibinfo{title}{Quenches near ising quantum criticality as a
  challenge for artificial neural networks}.
\newblock \emph{\bibinfo{journal}{Phys. Rev. B}} \textbf{\bibinfo{volume}{98}},
  \bibinfo{pages}{024311} (\bibinfo{year}{2018}).
\newblock \urlprefix\url{https://link.aps.org/doi/10.1103/PhysRevB.98.024311}.

\bibitem{amari_natural_1998}
\bibinfo{author}{Amari, S.~I.}
\newblock \bibinfo{title}{Natural gradient works efficiently in learning}.
\newblock \emph{\bibinfo{journal}{Neural Computation}}
  \textbf{\bibinfo{volume}{10}}, \bibinfo{pages}{251--276}
  (\bibinfo{year}{1998}).
\newblock
  \urlprefix\url{http://www.mitpressjournals.org/doi/10.1162/089976698300017746}.

\bibitem{Yuan2019}
\bibinfo{author}{Yuan, X.}, \bibinfo{author}{Endo, S.}, \bibinfo{author}{Zhao,
  Q.}, \bibinfo{author}{Li, Y.} \& \bibinfo{author}{Benjamin, S.~C.}
\newblock \bibinfo{title}{Theory of variational quantum simulation}.
\newblock \emph{\bibinfo{journal}{Quantum}} \textbf{\bibinfo{volume}{3}},
  \bibinfo{pages}{191} (\bibinfo{year}{2019}).
\newblock \urlprefix\url{https://quantum-journal.org/papers/q-2019-10-07-191/}.

\bibitem{stokes_quantum_2020}
\bibinfo{author}{Stokes, J.}, \bibinfo{author}{Izaac, J.},
  \bibinfo{author}{Killoran, N.} \& \bibinfo{author}{Carleo, G.}
\newblock \bibinfo{title}{Quantum natural gradient}.
\newblock \emph{\bibinfo{journal}{Quantum}} \textbf{\bibinfo{volume}{4}},
  \bibinfo{pages}{269} (\bibinfo{year}{2020}).
\newblock \urlprefix\url{https://quantum-journal.org/papers/q-2020-05-25-269/}.

\bibitem{Vogt2015}
\bibinfo{author}{Vogt, N.} \emph{et~al.}
\newblock \bibinfo{title}{{One-dimensional Josephson junction arrays: Lifting
  the Coulomb blockade by depinning}}.
\newblock \emph{\bibinfo{journal}{Physical Review B}}
  \textbf{\bibinfo{volume}{92}}, \bibinfo{pages}{045435}
  (\bibinfo{year}{2015}).
\newblock \urlprefix\url{http://dx.doi.org/10.1103/PhysRevB.92.045435}.

\bibitem{Martinoli2000}
\bibinfo{author}{Martinoli, P.} \& \bibinfo{author}{Leemann, C.}
\newblock \bibinfo{title}{{Two Dimensional Josephson Junction Arrays}}.
\newblock \emph{\bibinfo{journal}{Journal of Low Temperature Physics}}
  \textbf{\bibinfo{volume}{118}}, \bibinfo{pages}{699--731}
  (\bibinfo{year}{2000}).
\newblock
  \urlprefix\url{https://link.springer.com/article/10.1023/A:1004651730459}.

\bibitem{Kockum2019}
\bibinfo{author}{Kockum, A.~F.} \& \bibinfo{author}{Nori, F.}
\newblock \bibinfo{title}{Quantum bits with josephson junctions}.
\newblock \emph{\bibinfo{journal}{Springer Series in Materials Science}}
  \textbf{\bibinfo{volume}{286}}, \bibinfo{pages}{703--741}
  (\bibinfo{year}{2019}).
\newblock \urlprefix\url{http://dx.doi.org/10.1007/978-3-030-20726-7_17}.
\newblock \eprint{1908.09558}.

\bibitem{Jose1977}
\bibinfo{author}{José, J.~V.}, \bibinfo{author}{Kadanoff, L.~P.},
  \bibinfo{author}{Kirkpatrick, S.} \& \bibinfo{author}{Nelson, D.~R.}
\newblock \bibinfo{title}{Renormalization, vortices, and symmetry-breaking
  perturbations in the two-dimensional planar model}.
\newblock \emph{\bibinfo{journal}{Physical Review B}}
  \textbf{\bibinfo{volume}{16}}, \bibinfo{pages}{1217--1241}
  (\bibinfo{year}{1977}).
\newblock \urlprefix\url{https://link.aps.org/doi/10.1103/PhysRevB.16.1217}.

\bibitem{Stokes2021}
\bibinfo{author}{Stokes, J.}, \bibinfo{author}{De, S.},
  \bibinfo{author}{Veerapaneni, S.} \& \bibinfo{author}{Carleo, G.}
\newblock \bibinfo{title}{Continuous-variable neural-network quantum states and
  the quantum rotor model}  (\bibinfo{year}{2021}).
\newblock \urlprefix\url{http://arxiv.org/abs/2107.07105}.

\bibitem{Jiang2019}
\bibinfo{author}{Jiang, W.}, \bibinfo{author}{Pan, G.}, \bibinfo{author}{Liu,
  Y.} \& \bibinfo{author}{Meng, Z.~Y.}
\newblock \bibinfo{title}{Solving quantum rotor model with different monte
  carlo techniques}.
\newblock \emph{\bibinfo{journal}{Chinese Physics B}}
  \textbf{\bibinfo{volume}{31}} (\bibinfo{year}{2019}).
\newblock \urlprefix\url{http://arxiv.org/abs/1912.08229}.

\bibitem{trebst22}
\bibinfo{author}{Berke, C.}, \bibinfo{author}{Varvelis, E.},
  \bibinfo{author}{Trebst, S.}, \bibinfo{author}{Altland, A.} \&
  \bibinfo{author}{DiVincenzo, D.~P.}
\newblock \bibinfo{title}{Transmon platform for quantum computing challenged by
  chaotic fluctuations}.
\newblock \emph{\bibinfo{journal}{Nature Communications}}
  \textbf{\bibinfo{volume}{13}}, \bibinfo{pages}{2495} (\bibinfo{year}{2022}).
\newblock \urlprefix\url{https://doi.org/10.1038/s41467-022-29940-y}.

\bibitem{Becca2017}
\bibinfo{author}{Becca, F.} \& \bibinfo{author}{Sorella, S.}
\newblock \emph{\bibinfo{title}{Quantum Monte Carlo Approaches for Correlated
  Systems}} (\bibinfo{publisher}{Cambridge University Press},
  \bibinfo{year}{2017}).

\bibitem{Sorella1998}
\bibinfo{author}{Sorella, S.}
\newblock \bibinfo{title}{Green function monte carlo with stochastic
  reconfiguration}.
\newblock \emph{\bibinfo{journal}{Physical Review Letters}}
  \textbf{\bibinfo{volume}{80}}, \bibinfo{pages}{4558--4561}
  (\bibinfo{year}{1998}).
\newblock \urlprefix\url{https://link.aps.org/doi/10.1103/PhysRevLett.80.4558}.

\bibitem{Metropolis1953}
\bibinfo{author}{Metropolis, N.}, \bibinfo{author}{Rosenbluth, A.~W.},
  \bibinfo{author}{Rosenbluth, M.~N.}, \bibinfo{author}{Teller, A.~H.} \&
  \bibinfo{author}{Teller, E.}
\newblock \bibinfo{title}{Equation of state calculations by fast computing
  machines}.
\newblock \emph{\bibinfo{journal}{The Journal of Chemical Physics}}
  \textbf{\bibinfo{volume}{21}}, \bibinfo{pages}{1087--1092}
  (\bibinfo{year}{1953}).

\bibitem{Hastings1970}
\bibinfo{author}{Hastings, W.~K.}
\newblock \bibinfo{title}{Monte carlo sampling methods using markov chains and
  their applications}.
\newblock \emph{\bibinfo{journal}{Biometrika}} \textbf{\bibinfo{volume}{57}},
  \bibinfo{pages}{97--109} (\bibinfo{year}{1970}).

\bibitem{Bogacki1989}
\bibinfo{author}{Bogacki, P.} \& \bibinfo{author}{Shampine, L.}
\newblock \bibinfo{title}{A 3(2) pair of runge - kutta formulas}.
\newblock \emph{\bibinfo{journal}{Applied Mathematics Letters}}
  \textbf{\bibinfo{volume}{2}}, \bibinfo{pages}{321--325}
  (\bibinfo{year}{1989}).

\bibitem{Butcher2008}
\bibinfo{author}{Butcher, J.~C.}
\newblock \emph{\bibinfo{title}{Numerical Methods for Ordinary Differential
  Equations}} (\bibinfo{publisher}{John Wiley \& Sons, Ltd},
  \bibinfo{year}{2008}).

\bibitem{Press1992}
\bibinfo{author}{Press, W.~H.}, \bibinfo{author}{Teukolsky, S.~A.},
  \bibinfo{author}{Vetterling, W.~T.} \& \bibinfo{author}{Flannery, B.~P.}
\newblock \emph{\bibinfo{title}{Numerical Recipes in C}}
  (\bibinfo{publisher}{Cambridge University Press},
  \bibinfo{address}{Cambridge, USA}, \bibinfo{year}{1992}),
  \bibinfo{edition}{second} edn.

\bibitem{Neal2012}
\bibinfo{author}{Neal, R.~M.}
\newblock \emph{\bibinfo{title}{Handbook of Markov Chain Monte Carlo}}
  (\bibinfo{publisher}{Chapman and Hall/CRC}, \bibinfo{year}{2011}).
\newblock \urlprefix\url{https://www.taylorfrancis.com/books/9781420079425}.

\bibitem{Betancourt2017}
\bibinfo{author}{Betancourt, M.}
\newblock \bibinfo{title}{A conceptual introduction to hamiltonian monte carlo}
   (\bibinfo{year}{2017}).
\newblock \urlprefix\url{http://arxiv.org/abs/1701.02434}.

\bibitem{Nesterov2009}
\bibinfo{author}{Nesterov, Y.}
\newblock \bibinfo{title}{Primal-dual subgradient methods for convex problems}.
\newblock \emph{\bibinfo{journal}{Mathematical Programming}}
  \textbf{\bibinfo{volume}{120}}, \bibinfo{pages}{221--259}
  (\bibinfo{year}{2009}).
\newblock \urlprefix\url{http://link.springer.com/10.1007/s10107-007-0149-x}.

\bibitem{Hoffman2011}
\bibinfo{author}{Hoffman, M.~D.} \& \bibinfo{author}{Gelman, A.}
\newblock \bibinfo{title}{The no-u-turn sampler: Adaptively setting path
  lengths in hamiltonian monte carlo}.
\newblock \emph{\bibinfo{journal}{Journal of Machine Learning Research}}
  \textbf{\bibinfo{volume}{15}}, \bibinfo{pages}{1593--1623}
  (\bibinfo{year}{2011}).
\newblock \urlprefix\url{https://arxiv.org/abs/1111.4246v1}.

\bibitem{LeCun2015}
\bibinfo{author}{LeCun, Y.}, \bibinfo{author}{Bengio, Y.} \&
  \bibinfo{author}{Hinton, G.}
\newblock \bibinfo{title}{Deep learning.}
\newblock \emph{\bibinfo{journal}{Nature}} \textbf{\bibinfo{volume}{521}},
  \bibinfo{pages}{436--44} (\bibinfo{year}{2015}).
\newblock \urlprefix\url{http://www.ncbi.nlm.nih.gov/pubmed/26017442}.

\bibitem{Carleo2019}
\bibinfo{author}{Carleo, G.} \emph{et~al.}
\newblock \bibinfo{title}{Machine learning and the physical sciences}.
\newblock \emph{\bibinfo{journal}{Reviews of Modern Physics}}
  \textbf{\bibinfo{volume}{91}}, \bibinfo{pages}{045002}
  (\bibinfo{year}{2019}).
\newblock
  \urlprefix\url{https://link.aps.org/doi/10.1103/RevModPhys.91.045002}.

\bibitem{Pescia2022}
\bibinfo{author}{Pescia, G.}, \bibinfo{author}{Han, J.},
  \bibinfo{author}{Lovato, A.}, \bibinfo{author}{Lu, J.} \&
  \bibinfo{author}{Carleo, G.}
\newblock \bibinfo{title}{Neural-network quantum states for periodic systems in
  continuous space}.
\newblock \emph{\bibinfo{journal}{Physical Review Research}}
  \textbf{\bibinfo{volume}{4}}, \bibinfo{pages}{023138} (\bibinfo{year}{2022}).
\newblock
  \urlprefix\url{https://link.aps.org/doi/10.1103/PhysRevResearch.4.023138}.

\bibitem{heyl13}
\bibinfo{author}{Heyl, M.}, \bibinfo{author}{Polkovnikov, A.} \&
  \bibinfo{author}{Kehrein, S.}
\newblock \bibinfo{title}{Dynamical quantum phase transitions in the
  transverse-field ising model}.
\newblock \emph{\bibinfo{journal}{Phys. Rev. Lett.}}
  \textbf{\bibinfo{volume}{110}}, \bibinfo{pages}{135704}
  (\bibinfo{year}{2013}).
\newblock
  \urlprefix\url{https://link.aps.org/doi/10.1103/PhysRevLett.110.135704}.

\bibitem{schmittsels19}
\bibinfo{author}{Schmitt, M.}, \bibinfo{author}{Sels, D.},
  \bibinfo{author}{Kehrein, S.} \& \bibinfo{author}{Polkovnikov, A.}
\newblock \bibinfo{title}{Semiclassical echo dynamics in the sachdev-ye-kitaev
  model}.
\newblock \emph{\bibinfo{journal}{Phys. Rev. B}} \textbf{\bibinfo{volume}{99}},
  \bibinfo{pages}{134301} (\bibinfo{year}{2019}).
\newblock \urlprefix\url{https://link.aps.org/doi/10.1103/PhysRevB.99.134301}.

\bibitem{Medvidovic2021}
\bibinfo{author}{Medvidović, M.} \& \bibinfo{author}{Carleo, G.}
\newblock \bibinfo{title}{Classical variational simulation of the quantum
  approximate optimization algorithm}.
\newblock \emph{\bibinfo{journal}{npj Quantum Information}}
  \textbf{\bibinfo{volume}{7}}, \bibinfo{pages}{101} (\bibinfo{year}{2021}).
\newblock \urlprefix\url{https://www.nature.com/articles/s41534-021-00440-z}.

\bibitem{Jonsson2018}
\bibinfo{author}{Jónsson, B.}, \bibinfo{author}{Bauer, B.} \&
  \bibinfo{author}{Carleo, G.}
\newblock \bibinfo{title}{Neural-network states for the classical simulation of
  quantum computing} (\bibinfo{year}{2018}).
\newblock \urlprefix\url{http://arxiv.org/abs/1808.05232}.

\bibitem{Mandelstam1991}
\bibinfo{author}{Mandelstam, L.} \& \bibinfo{author}{Tamm, I.}
\newblock \emph{\bibinfo{title}{The Uncertainty Relation Between Energy and
  Time in Non-relativistic Quantum Mechanics}}, \bibinfo{pages}{115--123}
  (\bibinfo{publisher}{Springer Berlin Heidelberg}, \bibinfo{address}{Berlin,
  Heidelberg}, \bibinfo{year}{1991}).
\newblock \urlprefix\url{https://doi.org/10.1007/978-3-642-74626-0_8}.

\bibitem{Vidal2003}
\bibinfo{author}{Vidal, G.}
\newblock \bibinfo{title}{{Efficient classical simulation of slightly entangled
  quantum computations}}.
\newblock \emph{\bibinfo{journal}{Physical Review Letters}}
  \textbf{\bibinfo{volume}{91}}, \bibinfo{pages}{147902}
  (\bibinfo{year}{2003}).
\newblock
  \urlprefix\url{https://link.aps.org/doi/10.1103/PhysRevLett.91.147902}.
\newblock \eprint{0301063}.

\bibitem{Vidal2004}
\bibinfo{author}{Vidal, G.}
\newblock \bibinfo{title}{{Efficient simulation of one-dimensional quantum
  many-body systems}}.
\newblock \emph{\bibinfo{journal}{Physical Review Letters}}
  \textbf{\bibinfo{volume}{93}}, \bibinfo{pages}{040502}
  (\bibinfo{year}{2004}).
\newblock
  \urlprefix\url{https://journals.aps.org/prl/abstract/10.1103/PhysRevLett.93.040502}.
\newblock \eprint{0310089}.

\bibitem{Carleo2017}
\bibinfo{author}{Carleo, G.} \& \bibinfo{author}{Troyer, M.}
\newblock \bibinfo{title}{Solving the quantum many-body problem with artificial
  neural networks}.
\newblock \emph{\bibinfo{journal}{Science}} \textbf{\bibinfo{volume}{355}},
  \bibinfo{pages}{602--606} (\bibinfo{year}{2017}).

\bibitem{Haegeman2011}
\bibinfo{author}{Haegeman, J.} \emph{et~al.}
\newblock \bibinfo{title}{Time-dependent variational principle for quantum
  lattices}.
\newblock \emph{\bibinfo{journal}{Physical Review Letters}}
  \textbf{\bibinfo{volume}{107}}, \bibinfo{pages}{070601}
  (\bibinfo{year}{2011}).

\bibitem{Haegeman2016}
\bibinfo{author}{Haegeman, J.}, \bibinfo{author}{Lubich, C.},
  \bibinfo{author}{Oseledets, I.}, \bibinfo{author}{Vandereycken, B.} \&
  \bibinfo{author}{Verstraete, F.}
\newblock \bibinfo{title}{Unifying time evolution and optimization with matrix
  product states}.
\newblock \emph{\bibinfo{journal}{Physical Review B}}
  \textbf{\bibinfo{volume}{94}}, \bibinfo{pages}{165116}
  (\bibinfo{year}{2016}).

\bibitem{Verstraete2004}
\bibinfo{author}{Verstraete, F.} \& \bibinfo{author}{Cirac, J.~I.}
\newblock \bibinfo{title}{Renormalization algorithms for quantum-many body
  systems in two and higher dimensions}  (\bibinfo{year}{2004}).
\newblock \urlprefix\url{http://arxiv.org/abs/cond-mat/0407066}.

\bibitem{jax2018github}
\bibinfo{author}{Bradbury, J.} \emph{et~al.}
\newblock \bibinfo{title}{{JAX}: composable transformations of
  {P}ython+{N}um{P}y programs} (\bibinfo{year}{2018}).
\newblock \urlprefix\url{http://github.com/google/jax}.

\bibitem{flax2020github}
\bibinfo{author}{Heek, J.} \emph{et~al.}
\newblock \bibinfo{title}{{F}lax: A neural network library and ecosystem for
  {JAX}} (\bibinfo{year}{2020}).
\newblock \urlprefix\url{http://github.com/google/flax}.

\bibitem{numpy}
\bibinfo{author}{Harris, C.~R.} \emph{et~al.}
\newblock \bibinfo{title}{{Array programming with NumPy}}.
\newblock \emph{\bibinfo{journal}{Nature}} \textbf{\bibinfo{volume}{585}},
  \bibinfo{pages}{357--362} (\bibinfo{year}{2020}).
\newblock \eprint{2006.10256}.

\bibitem{scipy}
\bibinfo{author}{Virtanen, P.} \emph{et~al.}
\newblock \bibinfo{title}{Scipy 1.0: fundamental algorithms for scientific
  computing in python}.
\newblock \emph{\bibinfo{journal}{Nature Methods}}
  \textbf{\bibinfo{volume}{17}}, \bibinfo{pages}{261--272}
  (\bibinfo{year}{2020}).
\newblock \urlprefix\url{http://www.nature.com/articles/s41592-019-0686-2}.

\bibitem{matplotlib}
\bibinfo{author}{Hunter, J.~D.}
\newblock \bibinfo{title}{{Matplotlib: A 2D graphics environment}}.
\newblock \emph{\bibinfo{journal}{Comput. Sci. Eng.}}
  \textbf{\bibinfo{volume}{9}}, \bibinfo{pages}{99--104}
  (\bibinfo{year}{2007}).

\bibitem{Carpenter2017}
\bibinfo{author}{Carpenter, B.} \emph{et~al.}
\newblock \bibinfo{title}{Stan: A probabilistic programming language}.
\newblock \emph{\bibinfo{journal}{Journal of Statistical Software}}
  \textbf{\bibinfo{volume}{76}} (\bibinfo{year}{2017}).
\newblock \urlprefix\url{http://www.jstatsoft.org/v76/i01/}.

\bibitem{Butcher1963}
\bibinfo{author}{Butcher, J.~C.}
\newblock \bibinfo{title}{Coefficients for the study of runge-kutta integration
  processes}.
\newblock \emph{\bibinfo{journal}{Journal of the Australian Mathematical
  Society}} \textbf{\bibinfo{volume}{3}}, \bibinfo{pages}{185--201}
  (\bibinfo{year}{1963}).

\bibitem{White1992}
\bibinfo{author}{White, S.~R.}
\newblock \bibinfo{title}{{Density matrix formulation for quantum
  renormalization groups}}.
\newblock \emph{\bibinfo{journal}{Physical Review Letters}}
  \textbf{\bibinfo{volume}{69}}, \bibinfo{pages}{2863--2866}
  (\bibinfo{year}{1992}).
\newblock
  \urlprefix\url{https://journals.aps.org/prl/abstract/10.1103/PhysRevLett.69.2863}.

\end{thebibliography}

\onecolumngrid
\appendix

\section{Simulation details}
\label{appendix:tvmc_details}

In this Appendix, we mention some of the details of numerical simulations performed in this work that have not been discussed in the main text. We also clearly state different hyperparameters and their observed effect on performance and numerical stability.

\subsection{Hamiltonian Monte Carlo details}
\label{appendix:hmc}

As noted in the main text, the Hamiltonian Monte Carlo (HMC) algorithm used in this work has many important hyperparameters. To define the proposal, we must specify: the leapfrog integration length $L$, leapfrog step size $\varepsilon$ the mass matrix $M$. We fix $L$ heuristically and adaptively set $M$ and $\varepsilon$ during an extended \textit{warmup} phase for each Markov chain independently. We assume that the mass matrix is diagonal $M = \diag ( m_1, \ldots, m_N )$.

Before any samples are collected for evaluation of Eq.~\ref{eq:averages}, each chain is run for $N_\text{w}$ steps. Following the popular software package \textit{Stan}~\cite{Carpenter2017}, we subdivide the warmup period into $N_\text{p} + 2$ phases (\textit{windows}), each of which is one of two types:

\begin{itemize}
    \item \textbf{Fast}: Samples are collected and only step size $\varepsilon$ is adapted using the online optimization algorithm in Ref.~\cite{Nesterov2009}. Mass matrix remains unchanged. Fast windows are used to efficiently \textit{initialize} the chain by moving it towards a typical set of highly probable samples.
    \item \textbf{Slow}: Samples are collected and both step size $\varepsilon$ and the mass matrix $M$ are estimated. Step size is estimated the same way as in the fast window. Mass matrix elements are estimated as the variance of corresponding variables: $m_k = \Var (\theta _k)$ using the appropriate formula for the variance of periodic random variables presented in the main text, Sec.~\ref{sec:results}.
\end{itemize}

After initializing each $\theta _k \sim \text{Uniform}(-\pi, \pi)$, we begin the warmup phase with a single fast window of length $\nicefrac{N_\text{w}}{12}$, followed by five fast windows. The first fast window is $\nicefrac{N_\text{w}}{36}$ steps long with each subsequent slow window doubling in size. Finally, we end the warmup by running an additional fast window for the remaining $\nicefrac{N_\text{w}}{18}$ steps. After each window, the HMC transition kernel (the leapfrog ODE solver) is updated with adapted values for $\varepsilon$ and $M$ (for slow windows). After the final fast window, all hyperparameters are locked in and actual collection of the $N_\text{s}$ for Eq.~\ref{eq:averages} begins. The full list of relevant hyperparameters can be found in Table~\ref{tab:hmc_params}.

We use automatic differentiation (using JAX~\cite{jax2018github}) to obtain numerically exact gradients $ \nabla _{\ttheta} \ln p(\ttheta, t)$ of needed to run the leapfrog integrator. To avoid loss of accuracy or numerical instabilities through exponentiation, we employ the following identity:
\begin{equation}
    \ln p(\ttheta, t) = \ln \left\vert \psi _{\alpha (t)} (\ttheta) \right\vert ^2 = 2 \Re \left\lbrace \ln \psi _{\alpha (t)} (\ttheta) \right\rbrace \; ,
\end{equation}

\noindent when the logarithm of the wavefunction is parameterized instead of the wavefunction itself.

For completeness, we note that a common precaution against leapfrog integration getting stuck in regions of high curvature used in this work. Instead of fixing the integration length to a specific value $L=L_0$, it is randomly chosen between $(1-\gamma)L_0$ and $(1+\gamma)L_0$ each time the integrator is called, with a new hyperparameter $0 \leq \gamma < 1$. This \textit{jittering} of trajectory lengths can help HMC walkers move away from regions of high curvature if they get stuck~\cite{Betancourt2017, Neal2012, Carpenter2017}.

Finally, to collect more independent samples by utilizing modern massively parallel GPU hardware, we run $N_\text{c}$ such chains in parallel, each one warmed up independently.

Finally, we note that the HMC proposal outlined in Eq.~\ref{eq:leapfrog} approaches the RWM update:
\begin{equation}
    \ttheta ' = \ttheta + \sqrt{\Sigma} \, \vb{z} \; ; \quad \vb{z} \sim \mathcal{N} (0, \mathbbm{1}) \; ,
\end{equation}

\noindent in the limit of few leapfrog integrator steps: $L \rightarrow 1$. Indeed, for $L=1$ and small step sizes $\varepsilon$, Eq.~\ref{eq:leapfrog} becomes
\begin{equation}
    \ttheta ' = \ttheta (\epsilon) = \ttheta (0) + \varepsilon M^{-1} \ppi (\nicefrac{\varepsilon}{2}) = 
    \ttheta (0) + \varepsilon M^{-1} \ppi (0) - \frac{\varepsilon ^2}{2} M^{-1} \frac{\partial V}{\partial \ttheta } (\ttheta (0) ) = \ttheta (0) + \varepsilon M^{-1} \ppi (0) + \mathcal{O} (\varepsilon ^2) \; ,
\end{equation}

\noindent where $M^{-1}$ is equivalent in effect to the $\sqrt{\Sigma}$ matrix and $\ppi (0) \sim \mathcal{N} (0, M ^{-1})$ by construction in Eq.~\ref{eq:hmc_hamiltonian}.

\begin{table}[!t]
\label{tab:hmc_params}
\centering
\begin{tabular}{c|c|c|c|c}
    \textbf{Symbol} & \textbf{Name} & \textbf{Value} & \textbf{Domain} & \textbf{Description} \\
    \hline \hline
    $\varepsilon$ & Step size & \thead{Dynamically\\adapted} & $\mathbbm{R}$ & The leapfrog integrator step size. \\
    \hline
    $M$ & Mass matrix & \thead{Dynamically\\adapted} & $\mathbbm{R}^{N^2}$ & \thead{The covariance (metric) tensor of\\the dummy momentum variables $\ppi$.} \\
    \hline
    $L$ & (Average) integration length & 20 & $\mathbbm{N}$ & \thead{The number of leapfrog steps taken before\\proposing a sample. (If $\gamma > 0$, we relabel $L \rightarrow L_0$.)} \\
    \hline
    $\gamma$ & Jitter & 0.2 & $[0, 1 \rangle$ & \thead{Randomness for $L$ during sampling -- it is drawn\\uniformly between $(1-\gamma) L_0$ and $(1+\gamma)L_0$.} \\
    \hline
    $\varepsilon _0$ & Initial step size & 0.1 & $\mathbb{R}$ & \thead{A guess for the value of $\varepsilon$\\to refine during the warmup phase.} \\
    \hline
    $\delta$ & Target acceptance rate & 0.8 & $[0,1]$ & \thead{Target acceptance rate used for\\optimization of $\varepsilon$ by algorithm in Ref.~\cite{Nesterov2009}.} \\
    \hline
    $N_\text{w}$ & Length of warmup phase & 800 & $\mathbbm{N}$ & \thead{Total number of MC samples\\used for extended warmup.} \\
    \hline
    $N_\text{p}$ & Number of slow windows & 5 & $\mathbbm{N}$ & \thead{Total number of slow\\adaptation windows during warmup.} \\
    \hline
    $N_\text{s}$ & Number of samples & 2000 & $\mathbbm{N}$ & \thead{Total number of samples\\(per chain).} \\
    \hline
    $N_\text{c}$ & Number of chains & 20 & $\mathbbm{N}$ & \thead{Total number of independent\\Markov chains.} \\
\end{tabular}
\caption{
    The list of relevant hyperparameters for the Hamiltonian Monte Carlo algorithm with their values used in this work.
}
\end{table}

\subsection{Numerical regularization schemes}
\label{appendix:regularization}

After evaluating the averages in Eq.~\ref{eq:averages} at time $t$, one needs to solve the linear system $i\, S \dot{\alpha} = g$ to obtain $\dot{\alpha}$ needed to progress to time $t + \delta t$. Since the $S$ matrix is singular in most cases of interest, a robust regularization scheme is needed. As pointed out in the main text, replacing $S \rightarrow S + \epsilon \mathbbm{1}$ is often enough in the case of ground-state searches (imaginary-time evolution). We remark that this is equivalent to the L2-regularized least-squares solution of $i \, L \dot{\alpha} = h$.
\begin{equation}
    \dot{\alpha} = \argmin _{\dot{\alpha} \in \mathbbm{C}^P} \left\lbrace \norm{i \, L \, \dot{\alpha} - h} ^2 _2 + \epsilon \norm{\dot{\alpha}} ^2 _2 \right\rbrace
\end{equation}

\noindent where $L^\dagger L = S$ is the Cholesky decomposition of the $S$~matrix (assuming S is positive-definite), $L^\dagger h = g$, and $\norm{\cdot} _2$ is the standard euclidean 2-norm on $\mathbbm{C}^P$.

As outlined in the main text, we instead adopt a regularization scheme based on the spectrum of the $S$~matrix, $S = U \Sigma  U^\dagger$, where $\Sigma = \diag (\sigma ^2 _1, \ldots, \sigma ^2 _P)$. Our definition of the pseudoinverse is $S ^{-1} \approx U \Tilde{\Sigma} ^{-1} U^\dagger$ with
\begin{equation}
\label{eq:reg_details}
    \Tilde{\Sigma} ^{-1} _{\mu \nu} = f (\sigma ^2 _\mu) \; \frac{\delta _{\mu \nu} }{\sigma ^2 _\mu} \qquad \text{and} \qquad f(\sigma ^2) = \frac{1} { 1 + \left( \nicefrac{\lambda ^2}{\sigma ^2} \right) ^6 } \; .
\end{equation}

In the limit of $\lambda ^2 \rightarrow 0$, we recover the actual matrix inverse. As opposed to the more traditional choice of the step function $f(\sigma^2) = \theta (\sigma ^2 - \lambda^2)$, we find that choosing a smooth functional form for $f(\sigma^2)$ in Eq.~\ref{eq:reg_details} makes the adaptive time stepping in the top-level integration routine (see Appendix~\ref{appendix:tdvp_integrators}) more stable.

As noted in the main text, we set $\lambda ^2$ to:
\begin{equation}
    \lambda ^2 = \lambda ^2 (\sigma ^2 _1, \ldots, \sigma ^2 _P ) = \max \left( a_c, \; r_c \times \max _\mu (\sigma ^2 _\mu) \right) \; ,
\end{equation}

\noindent each iteration, with $a_c = 10^{-4}$ and $r_c = 10^{-2}$ chosen for 2D calculations and $a_c = 10^{-5}$ and $r_c = 10^{-4}$ for 1D benchmarks. To track potential over-regularization and as a measure of ansatz expressivity, we define the \textit{effective rank} $\rho (S) = \sum _\mu f(\sigma ^2 _\mu)$. Intuitively, since $0 < f(\sigma ^2) < 1$ for all eigenvalues $\sigma ^2$, $\rho (S)$ can be interpreted as the effective number of eigenvalues that have not been set to zero by the regularization function $f$. In other words, it corresponds to the number of parameters in $\alpha$ that get updated at time $t$.

We plot $\rho (S)$ as a function of time on Fig.~\ref{fig:details} for some simulated quenches. In all cases, we see that the effective rank increases rapidly to $\rho \sim 1$ at intermediate times that, for larger quenches, correspond to rapid oscillations and onset of vorticity. In those cases, it is natural to interpret this regime as almost all parameters $\alpha$ being important to capture the relevant physics. At later times, $\rho$ converges to values below $10\%$, indicating equilibration and less oscillatory behavior.

For completeness, we mention that alternative regularization techniques have been explored as well. For example, the method of Schmitt and Heyl in Ref.~\cite{Schmitt2020}, based on the signal-to-noise ratio (SNR) for each eigenvalue in $\sigma ^2$, represents a computationally and physically well-motivated approach. However, it did not bring any measurable performance improvement in our case.

\subsection{The time-dependent variational principle and ODE integrators}
\label{appendix:tdvp_integrators}

\begin{figure*}[t]
    \centering
    \includegraphics[width=\linewidth]{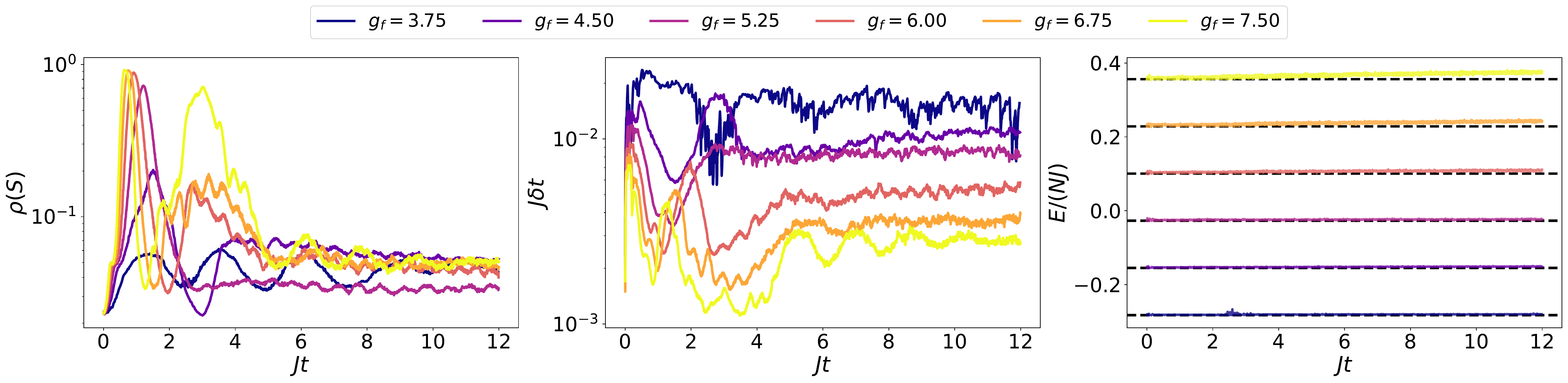}
    \caption{
        Some ODE integrator and regularization details.
        \textbf{Left}: The \textit{effective rank} of the $S$-matrix $\rho(S)$ as a function of time, reflecting the internal dimensionality of the parameter manifold $\alpha$, as discussed in Sec.~\ref{appendix:regularization}.
        \textbf{Center}: Time steps $\delta t$ taken by the adaptive ODE integrator of choice.
        \textbf{Right}: Approximate conservation of energy as a function of time, for different quenches.
    }
    \label{fig:details}
\end{figure*}

To make use of the TDVP action in Eq.~\ref{eq:tdvp} in the main text, to propagate the variational parameters forward in time, one must construct the corresponding Euler-Lagrange equations. To this end, we first manipulate the action into a more transparent form:
\begin{align}
\label{eq:tdvp_details}    
    \delta \mathcal{C}[\alpha] =& \; \delta \int \dd t \bra{\Psi _{\alpha (t)}} \left( i \frac{\dd}{\dd t} - H \right) \ket{\Psi _{\alpha (t)}} = \\
    =& \; \delta \int \dd t \left\lbrace \frac{i}{2} \frac{\braket{\psi _{\alpha (t)}}{\dot{\psi} _{\alpha (t)}} - \braket{\dot{\psi} _{\alpha (t)}}{\psi _{\alpha (t)}}}{\braket{\psi _{\alpha (t)}}} - \frac{\bra{\psi _{\alpha (t)}} H \ket{\psi _{\alpha (t)}}}{\braket{\psi _{\alpha (t)}}} \right\rbrace = \\
    =& \; - \delta \int \dd t \int \dd \ttheta \; \frac{ \left| \braket{\ttheta}{ \psi _{\alpha (t)}} \right| ^2}{ \braket{ \psi _{\alpha (t)}} } \left\lbrace \Im \sum _\mu \OO _\mu (\ttheta, t) \dot{\alpha} _\mu (t) + E_L (\ttheta, t) \right\rbrace \propto \label{eq:tdvp_details_3} \\
    \propto& \; \int \dd t \, \sum _\mu \left\lbrace i \sum _\nu S_{\mu \nu} (t) \, \dot{\alpha} _\nu (t) - g _\mu (t) \right\rbrace \delta \alpha _\mu (t) - \text{c.c.} \label{eq:tdvp_details_4} \; ,
\end{align}

\noindent where we used definitions of $S _{\mu \nu}$, $g_\mu$ and $E_L$ from Eqs.~\ref{eq:local_energy}~and~\ref{eq:averages}, respectively, as well as $\dot{\alpha} = \nicefrac{\dd \alpha}{\dd t}$. The explicit form of log-derivative operators $\OO _\mu$ introduced in the main text is
\begin{equation}
    \OO _\mu (\ttheta, t) = \frac{\partial}{\partial \alpha _\mu}  \ln \psi _{\alpha (t)} (\ttheta)
\end{equation}

\noindent in the $\ket{\ttheta}$ basis. We note that the expression in Eq.~\ref{eq:tdvp_details_3} and the definitions of $S$ and $g$ in Eq.~\ref{eq:tdvp_details_4} change form if $\psi _\alpha (\ttheta)$ cannot be interpreted as a holomorphic function of $\alpha$. The reader is referred to Ref.~\cite{Yuan2019} for detailed derivations.

We implement and experiment with a number of different Runge-Kutta~\cite{Press1992} (RK) ODE solvers. Heuristically, we notice that higher-order adaptive embedded solvers do help offset the effects of imperfect Monte Carlo estimates of $S$ and $g$ in two ways:

\begin{itemize}
    \item Adaptive solvers are naturally higher-order because of an embedded lower-order method. Using an adaptive solver can locally adjust the time step $\delta t$, usually drastically reducing the overall number of time steps required.
    \item The solution at $t + \delta t$ is constructed as a linear combination of solutions estimated on a fixed set of points within the interval $[t, t+\delta t]$. Any leftover errors in these intermediate estimates have a higher probability of canceling out.
\end{itemize}

In this work, we choose the adaptive third-order method with an embedded second order method using the Bogacki-Shampine~\cite{Bogacki1989, Butcher1963, Butcher2008, Press1992} pair of formulas. It balances being low-enough order to avoid wasting computational resources with still being high-enough order to allow for adaptive time-stepping.

For quenches shown in Fig.~\ref{fig:observables}, we show variations in $\delta t$ in Fig.~\ref{fig:details} (center panel). We note that the time stepping varies more in regions of higher curvature (kinetic energy, shorter times) and for larger quenches, successfully adjusting to conserve energy in all cases (Fig.~\ref{fig:details} right). For longer times, on the order of thermalization in observed quenches, variations are reduced and $\delta t$ approximately converges to a constant value.

\section{Tensor-network calculations}
\label{appendix:tensors}

In this Appendix we present the details of tensor-network calculations performed in this work. First, we lay out some of the conventions and formalism involved with treating the QRM in the discrete eigenbasis of the angular momentum operator $L_k$. This basis is useful for attacking the model with tensor-network methods or perturbation theory. Finally, we report the dependence of results reported in the main text on two different cutoffs -- one in singular value magnitude and one in local basis size.

\subsection{Angular momentum basis}
\label{subsection:tensors_basis}

In the $\ket{\ttheta}$ basis, we have $L_k = -i \, \partial _k$ (where we adopt the convention $\partial _k \equiv \partial _{\theta _k}$). Therefore:
\begin{equation}
    -i \frac{\partial}{\partial \theta} \braket{\theta}{m} = m \braket{\theta}{m} \qquad \Longrightarrow \qquad \braket{\theta}{m} = \frac{e^{-i m \theta}}{\sqrt{2 \pi}} \; ; \quad m \in \mathbb{Z} \; ,
\end{equation}

\noindent which is identical to eigenfunctions for a particle on a circle at each lattice site -- we have a product basis basis $\ket{\vb{m}} = \ket{m_1, \ldots, m_N}$. The Hamiltonian given in Eq.~\ref{eq:hamiltonian} then reads:
\begin{equation}
    \label{eq:ang_mom}
    \bra{\vb{m}'} H \ket{\vb{m}} = \frac{g J}{2} \sum _k m _k ^2 - J \sum _{\langle k, l \rangle} \bra{\vb{m}'} \hat{\vb{n}} _k \cdot \hat{\vb{n}} _l \ket{\vb{m}} \; .
\end{equation}

After inserting the identity $\mathbbm{1} = \int  \dd \ttheta \ketbra{\ttheta}$ into the second term and simple integration, we obtain
\begin{equation}
    \label{eq:potential_mat_el}
    \bra{m'_k, m'_l} \hat{\vb{n}} _k \cdot \hat{\vb{n}} _l \ket{m_k, m_l} = \frac{1}{2} \left( \delta _{m'_k,m_k+1} \delta _{m'_l,m_l-1} + \delta _{m'_k,m_k-1} \delta _{m'_l,m_l+1} \right) \; .
\end{equation}

\noindent where $\delta _{\cdot \cdot}$ is the Kronecker $\delta$ symbol. The structure of Eq.~\ref{eq:potential_mat_el} suggests rewriting the original Hamiltonian as
\begin{equation}
    H = \frac{g J}{2} \sum _k L ^2 _k - \frac{J}{2} \sum _{\langle k, l \rangle} \left( L^+ _k L^-_l + L^+ _l L^-_k \right)
\end{equation}

\noindent where
\begin{equation}
\label{eq:ladder_ops}
    L^+ _k \equiv \sum _{m_k \in \mathbbm{Z}} \ketbra{m_k +1}{m_k} \; \qquad \text{and} \qquad L^- _k \equiv \sum _{m_k \in \mathbbm{Z}} \ketbra{m_k - 1}{m_k} \; \qquad \text{so that} \qquad \comm{L^+_k}{L^-_k} = 0 \; ,
\end{equation}

\noindent for all $k \in \Lambda$. To perform tensor-network calculations, we truncate local basis states to $\left\lbrace \ket{-M}, \ldots, \ket{M} \right\rbrace $ so that $L^+ \ket{M} = L^- \ket{-M} = 0$. Namely, we set
\begin{equation}
    L^+ \rightarrow
    \begin{pmatrix}
    0 & 1 &   &        &   &   \\
      & 0 & 1 &        &   &   \\
      &   &   & \ddots &   &   \\
      &   &   &        & 0 & 1 \\
      &   &   &        &   & 0 \\
    \end{pmatrix} \; , \qquad
    L^- \rightarrow
    \begin{pmatrix}
    0 &   &   &        &   &   \\
    1 & 0 &   &        &   &   \\
      & 1 &   &        &   &   \\
      &   &   & \ddots & 0 &   \\
      &   &   &        & 1 & 0 \\
    \end{pmatrix}
\end{equation}

\noindent to build a matrix product operator (MPO) representation of the Hamiltonian. We use $M=5$ throughout. For use in real-time evolution through the TEBD algorithm in Sec.~\ref{sec:comparison} in the main text, we initialize the trial wavefunction as a matrix product state (MPS) \cite{White1992} (see Appendix~\ref{appendix:mps_ansatz}). We exploit
\begin{equation}
\label{eq:initial_state}
    \ket{\psi (t=0)} \propto \int \dd \ttheta \ket{\ttheta} = \prod _{i \in \Lambda } \ket{m_i = 0}
\end{equation}

\noindent to match the initial state given in the main text in Sec.~\ref{sec:comparison} by initializing the corresponding MPS to have bond dimension $\chi = 1$.

\begin{figure*}[t]
    \centering
    \includegraphics[width=\linewidth]{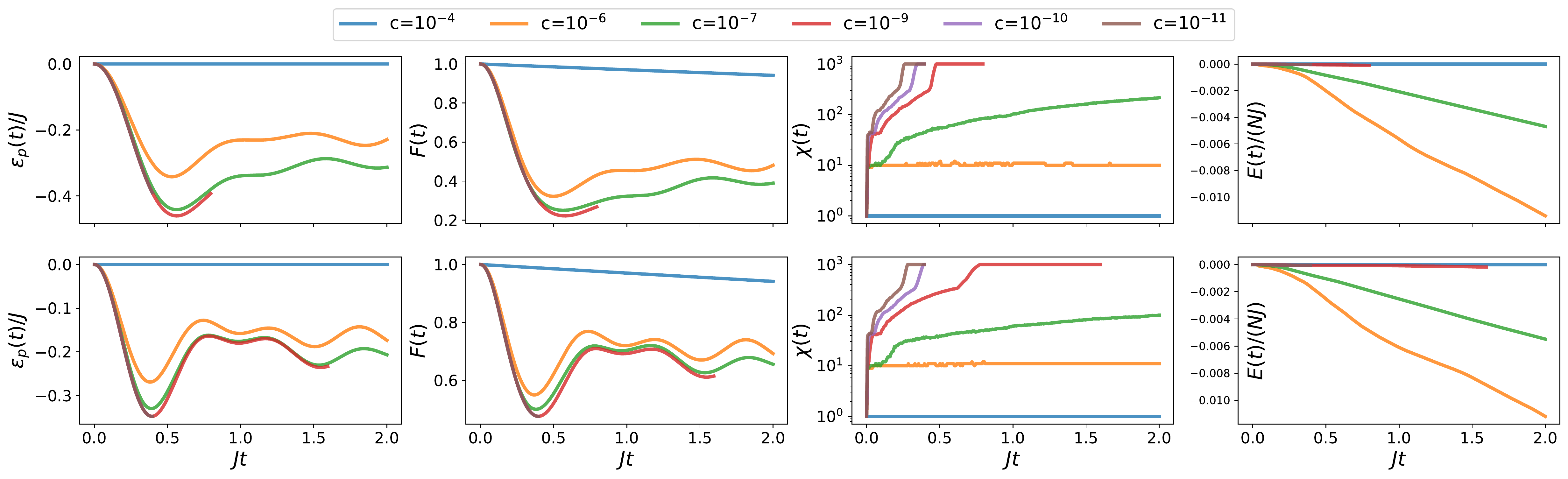}
    \caption{
        TEBD results for a range of different values of singular value cutoffs $c$. We plot time evolution of potential energy density $\epsilon _\text{p}$, fidelity $F$ as well as bond dimension $\chi$ growth. Additionally, we show energy conservation (or lack thereof). The initial state is set to the one described in Eq.~\ref{eq:initial_state} in all cases. Two independent sets of calculations were performed under the QRM Hamiltonian (Eq.~\ref{eq:hamiltonian}) with $g=5.6$ (top row) and $g=8.0$ (bottom row).
    }
    \label{fig:cutoffs}
\end{figure*}

\begin{figure*}[h!]
    \centering
    \includegraphics[width=\linewidth]{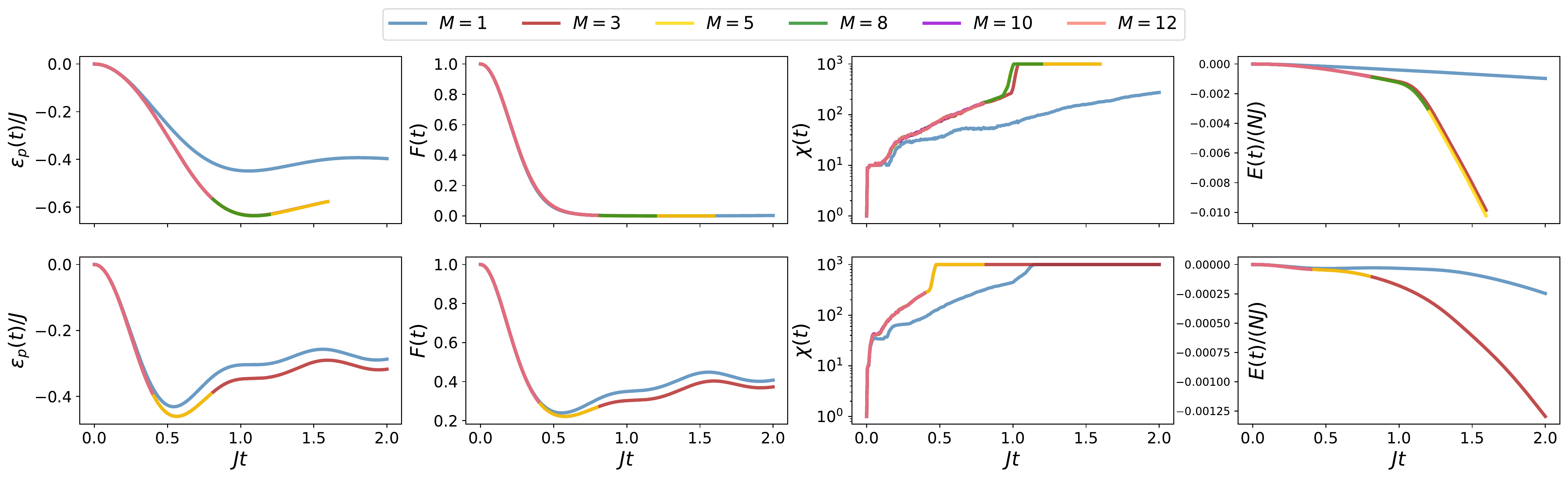}
    \caption{
        TEBD results for a range of different choices of truncated local bases ($M$). We plot the same observables for the same initial state as in Fig.~\ref{fig:cutoffs}. Similarly, two independent sets of calculations were performed with $g=2.0$ and $c=10^{-7}$ (top row) as well as $g=5.6$ and $c=10^{-9}$ (bottom row).
    }
    \label{fig:truncations}
\end{figure*}

\begin{figure*}[h!]
    \centering
    \includegraphics[width=\linewidth]{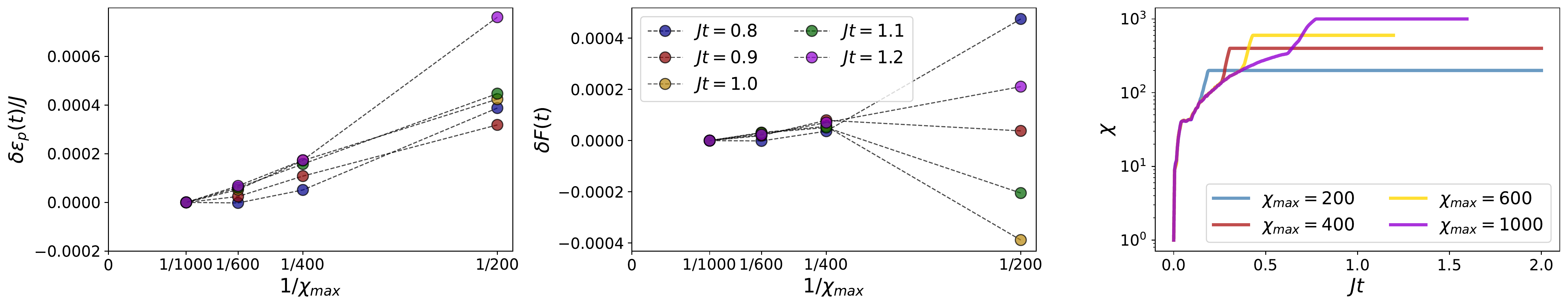}
    \caption{
        Convergence of TEBD calculations with respect to the cutoff bond dimension value $\chi _\text{max}$. We perform independent calculations for a fixed set of times, plot reference observable values obtained at different saturated bond dimensions. Plotted errors are defined as the difference between the value at bond dimension $\chi$ and the value at bond dimension $\chi = 1000$: $\delta A = A(\chi) - A(\chi = 1000) $ where $A=\epsilon _p$ in the left panel and $A = F$ in the middle panel.
    }
    \label{fig:chimax}
\end{figure*}

\subsection{Convergence and cutoffs}
\label{subsection:tensors_convergence}

To reinforce TEBD results as a benchmark in two dimensions, we further study the dependence of obtained results on the singular value cutoff $c$ and the local basis truncation parameter $M$ defined in subsection~\ref{subsection:tensors_basis}. In both cases, we perform a series of calculations on a $4 \times 4$ system for a set of different cutoffs and for two different values of the coupling constant $g$ in the QRM Hamiltonian (Eq.~\ref{eq:hamiltonian}).

In the case of the singular value cutoff $c$, we look at a range of values between $10^{-4}$ and $10^{-11}$ in Fig.~\ref{fig:cutoffs}. We observe that plotted curves seem to converge only for $c \leq 10^{-9} $. Satisfactory energy conservation is reached for $c \approx 10^{-9} $ as well as convergence of measured observables, except for the highest value $c=10^{-4}$. In that case, almost all singular values are discarded, extinguishing any nontrivial dynamics. Any further decrease in $c$ has the undesirable side-effect of rapidly growing the MPS bond dimension $\chi$, making longer simulations prohibitively expensive.

We explore a range of values for the local basis truncation parameter $M$ as well. For both reference values of the coupling constant $g$, we observe fast convergence towards self-consistent time evolution. These results indicate lower sensitivity to values of $M$, as long they lie above a threshold of $M \gtrsim 3$ ($M=5$ was used in the main text). Therefore, for short to intermediate time evolution, $J t \sim 1 - 10$, we expect the numerical cost of increased bond dimensions to remain dominated by singular value cutoff $c$.

In all cases, bond dimensions $\chi$ increase quickly and saturate at the cutoff value $\chi _\text{max}$, as is characteristic of MPS-based calculations in two dimensions. Therefore, to estimate any additional errors coming from the choice of $\chi _\text{max}$ (set to $10^3$ in the main text), we perform an additional set of independent calculations for a range of values, at $g=8.0$. Results are presented in Fig.~\ref{fig:chimax} where we plot estimated errors in reference observables. We discover that the relevant errors can be neglected for values as low as $\chi _\text{max} \approx 600$, at least for intermediate times. Reference observables quickly converge to the desired precision in this case.

\section{Variational Quantum States on a circle}
\label{appendix:ansatze}

In order to perform numerically efficient t-VMC iterations described in Sec.~\ref{sec:variational_sim}, we need to keep the parameter count in $\psi _\alpha$ relatively low (order $10^2$-$10^4$ on modern GPU hardware.). The reason for this constraint is the regularization scheme we employ to stabilize the QGT inverse in Eq.~\ref{eq:averages} -- each iteration requires us to diagonalize the $P \times P$ hermitian matrix $S$ which becomes prohibitively expensive and memory consuming for large parameter counts $P$. We note that iterative solvers such as the \textit{conjugate gradient} can formally help push the limit of tractable $P$ by several orders of magnitude. However, that speedup comes at the cost of having to rely on weaker regularization schemes that do not require the full QGT spectrum. In this Appendix, we describe different ansatzes (trial wavefunctions) considered in our simulations.

\subsection{The Jastrow wavefunction}
\label{appendix:jastrow}

The Jastrow wavefunction is defined as:
\begin{equation}
    \ln \psi _\alpha (\ttheta) = \frac{1}{2} \sum _{i j} w _{i j} \, \hat{\vb{n}} _i \cdot \hat{\vb{n}} _j = \frac{1}{2} \sum _{i j} w _{i j} \cos (\theta _i - \theta _j)
\end{equation}

\noindent where $\alpha = \{ w_{i j} \}$ is a symmetric matrix of variational parameters. The advantage of a simple Jastrow ansatz is the fact that the QGT given in Eq.~\ref{eq:averages} is never ill conditioned. However, Jastrow expressivity is limited compared to deeper neural-network quantum state parametrizations.

\subsection{Matrix Product States}
\label{appendix:mps_ansatz}

Using notation and conventions laid out in Appendix~\ref{appendix:tensors}, one can write down a traditional MPS ansatz in the discrete angular momentum basis $\ket{\vb{m}}$:
\begin{equation}
    \psi _\alpha (\ttheta ) = \sum ^M _{m_1 = -M} \cdots \sum ^M _{m_N = -M} c_{m_1 \cdots m_N} \; e^{ -i \sum _i m_i \theta _i } \; ; \quad
    c_{m_1 \cdots m_N} = \sum ^{\chi_1} _{l_{1} = 1} \cdots \sum ^{\chi_N} _{l_{N} = 1} A^{m_1} _{l_1} A^{m_2} _{l_1 l_2} \cdots A^{m_{N-1}} _{l_{N-2} l_{N-1}} A^{m_N} _{l_{N-1}} \; .
\end{equation}

In one spatial dimension and for ground-state searches (imaginary time evolution), this trial wavefunction form is the most accurate due to the density matrix renormalization group algorithm (DMRG)~\cite{White1992} algorithm. Expressivity is controlled by \textit{bond dimensions} $\chi_i$ and the basis truncation parameter $M$.

\subsection{The (circular) restricted Boltzmann machine}
\label{appendix:rbm}

The circular restricted Boltzman machine (RBM)~\cite{Stokes2021} is defined as:
\begin{equation}
    \label{eq:rbm}
    \psi _\alpha (\ttheta) \propto \int \dd \mu (\hat{\vb{h}}) \; \exp \left\{ \sum _j \vb{a}_j \cdot \hat{\vb{n}} _j + \sum _k \vb{b} _k \cdot \hat{\vb{h}} _k  + \sum _{j k} w_{j k} \, \hat{\vb{n}} _j \cdot \hat{\vb{h}} _k \right\}
\end{equation}

\noindent where $\alpha = \{ \vb{a}_j, \vb{b} _k, w_{jk} \}$ are variational parameters and $\dd \mu (\hat{\vb{h}})$ is the relevant measure for \textit{hidden units} $\hat{\vb{h}} _k$. It is natural to choose hidden units to have the same intrinsic Hilbert space as visible rotors $\hat{\vb{n}}_j$. Therefore, for the $O(2)$ quantum rotor model, we choose
\begin{equation}
    \hat{\vb{h}} _k = (\cos \phi _k, \sin \phi _k ) \qquad \text{so that} \qquad \dd \mu (\hat{\vb{h}}) = \dd \bm{\phi} = \dd \phi _i \, \dd \phi _2 \, \cdots \, \dd \phi _{N_\text{h}} \; ,
\end{equation}

\noindent up to an overall multiplicative constant. We note that the number hidden units $N_\text{h}$ is a hyperparameter and can be increased to control ansatz expressivity.

After performing the integrals in Eq.~\ref{eq:rbm}, one obtains the following closed-form expression:
\begin{equation}
    \ln \psi _\alpha (\ttheta) =  \sum _{j=1} ^N \vb{a}_j \cdot \hat{\vb{n}} _j + \sum _{k=1} ^{N_\text{h}} \ln I _0 \left( \sqrt { \sum\nolimits _l \left( \vb{x} _k \right) _l ^2} \right)
\end{equation}

\noindent where $(\vb{x} _k) _l$ stands for the $l$~th component of vector $\vb{x}_k = \vb{b} _k + \sum _j w_{j k} \hat{\vb{n}} _j $ and $I_0$ is the zeroth-order modified Bessel function of the first kind.

Instead of a full dense matrix, we can restrict the general linear map $\hat{\vb{n}} _j \mapsto \sum _j w_{j k} \hat{\vb{n}} _j$ to a convolution, assuming that underlying rotors are arranged in a square lattice. This restriction cuts $P$ down by approximately an order of magnitude while not sacrificing any measurable accuracy in ground state optimization tasks.

\subsection{Activation functions}
\label{appendix:activations}

In order to define an analytic ansatz $\psi _\alpha$ with no hidden sigularities, care must be taken when choosing activation functions for complex-valued inputs. Informally, singularities often appear in one of the following two ways, when using holomorphic activations:

\begin{itemize}
    \item A well-behaved function (or its derivatives) on the real axis has singularities on the imaginary axis. This is the case for $\tanh$ and $\ln I_0$ from Eq.~\ref{eq:rbm}, for example.
    \item An otherwise well-behaved function has a branch cut that is crossed during time evolution. Side effects include sudden jumps in conserved quantities during real-time evolution. This is the case for $\ln I_0$ from Eq.~\ref{eq:rbm} and similar functions involving logarithms and/or roots.
\end{itemize}

There are two solutions to this problem. As noted in the main text, one can restrict themselves to (higher-order) polynomial activations which are analytic everywhere and have no branch cuts. Inspired by Eq.~\ref{eq:rbm}, we use Taylor expansions of $\ln I_0$ and its gradient:
\begin{equation}
    \ln I_0(z) = \frac{z^2}{4} - \frac{z^4}{64} + \frac{z^6}{576} + \mathcal{O}(z^8)
    \quad \text{and} \quad
    \frac{I_1(z)}{I_0(z)} = \frac{z}{2} - \frac{z^3}{16} + \frac{z^5}{96} + \mathcal{O}(z^7) \; .
\end{equation}

\noindent This approach has the advantage of maintaining the holomorphic dependence of $\psi _\alpha$ on $\alpha$ and preserving the form of Eq.~\ref{eq:averages}. We note that if the effect of isolated singularities is not as important, Padé approximants often provide better approximations of target functions (and better ground-state energies) while still eliminating branch cuts.

The second option is abandoning holomorphicity in parameters $\alpha$ and applying well-behaved real activations to real and imaginary parts of the input separately. In that case, Eq.~\ref{eq:averages} must be corrected. We refer interested readers to the excellent overview of subtleties associated with complex parameters in Ref.~\cite{Yuan2019}.

\section{The $R^2$ performance metric}
\label{appendix:performance}

In the main text, we define a measure of time-dependent integration error in Eq~\ref{eq:full_residual}, following Refs.~\cite{Schmitt2020, Carleo2017}. The Fubini-Study distance $\mathcal{D}(\cdot, \cdot)$ on the Hilbert space $\mathcal{H}$ defined as
\begin{equation}
    \mathcal{D}(\psi, \phi) = \cos ^{-1} \left( \sqrt{F(\psi, \phi)} \right) = \cos ^{-1} \left( \sqrt{\frac{\left| \braket{\psi}{\phi} \right| ^2}{\braket{\psi} \braket{\phi}}} \right) \, .
\end{equation}

Using a consistent Taylor expansion in the limit of $\delta t \ll J ^{-1}$,
\begin{align}
    e^{-i H \delta t} =& \mathbbm{1} - i H \delta t + \mathcal{O}(\delta t ^2) \\
    \ket{\psi (t + \delta t)} =& \left( 1 + \delta t \sum _\mu \dot{\alpha} _\mu \OO _\mu\right) \ket{\psi _{\alpha (t)}}  + \mathcal{O}(\delta t ^2) \; ,
\end{align}

\noindent authors in Ref.~\cite{Schmitt2020} rewrite Eq.~\ref{eq:full_residual} as:
\begin{equation}
\label{eq:residual_expansion}
    r^2 (t) = 1 - \frac{1}{\Var _t H} \sum _{\mu \nu} S^{-1} _{\mu \nu} g^* _{\mu} g_{\nu} + \mathcal{O}(\delta t ^2) \; .
\end{equation}

In Eq.~\ref{eq:residual_expansion}, we used notation from Eq.~\ref{eq:averages} in the main text with
\begin{equation}
    \Var _t H = \avg{H^2} _t - \avg{H} _t ^2 \approx \avg{ \left| E_L - \avg{E _L} \right| ^2} _t
\end{equation}

Finally, the $R^2$ figure of merit is constructed as a time integral of $r^2$: $R^2 (t) = \int _0 ^t r^2(s) \dd s$. We note that $r^2 (t)$ and $R^2 (t)$ are readily available for estimation through Monte Carlo sampling.

\end{document}